\documentclass[aps,prd,groupedaddress,nofootinbib,amssymb,eqsecnum,epsfig]{revtex4}
\pdfoutput=1
\usepackage{graphicx}
\usepackage{bm}
\usepackage{amsmath}
\usepackage{hyperref}

\def \lleq {\lower0.9ex\hbox{ $\buildrel < \over \sim$} ~}
\def \ggeq {\lower0.9ex\hbox{ $\buildrel > \over \sim$} ~}

\def \beq  {\begin{equation}}
\def \eeq  {\end{equation}}
\def \ber  {\begin{eqnarray}}
\def \eer  {\end{eqnarray}}
\def \erf {\rm erf}
\def \tr {\rm tr}

\begin{document}
\newcommand{\newc}{\newcommand}

\newc{\be}{\begin{equation}}
\newc{\ee}{\end{equation}}
\newc{\ba}{\begin{eqnarray}}
\newc{\ea}{\end{eqnarray}}
\newc{\bea}{\begin{eqnarray*}}
\newc{\eea}{\end{eqnarray*}}
\newc{\D}{\partial}
\newc{\ie}{{\it i.e.} }
\newc{\eg}{{\it e.g. } }
\newc{\etc}{{\it etc.} }
\newc{\etal}{{\it et al.}}
\newcommand{\nn}{\nonumber}
\newc{\ra}{\rightarrow}
\newc{\lra}{\leftrightarrow}
\newc{\lsim}{\buildrel{<}\over{\sim}}
\newc{\gsim}{\buildrel{>}\over{\sim}}

\newcommand{\JGB}[1]{\textcolor{red}{\bf JGB: #1}}

\title{Is the Jeffreys' scale a reliable tool for Bayesian model comparison in cosmology?}
\author{Savvas Nesseris}
\author{Juan Garc\'ia-Bellido}
\email{savvas.nesseris@uam.es, juan.garciabellido@uam.es}
\affiliation{Instituto de F\'isica Te\'orica UAM-CSIC, Universidad Auton\'oma de Madrid,
Cantoblanco, 28049 Madrid, Spain}

\date{\today}

\begin{abstract}
We are entering an era where progress in cosmology is driven by data, and alternative models will have to be compared and ruled out according to some consistent criterium. The most conservative and widely used approach is Bayesian model comparison. In this paper we explicitly calculate the Bayes factors for all models that are linear with respect to their parameters. We do this in order to test the so called Jeffreys' scale and determine analytically how accurate its predictions are in a simple case where we fully understand and can calculate everything analytically. We also discuss the case of nested models, e.g. one with $M_1$ and another with $M_2\supset M_1$ parameters and we derive analytic expressions for both the Bayes factor and the Figure of Merit, defined as the inverse area of the model parameter's confidence contours. With all this machinery and the use of an explicit example we demonstrate that the threshold nature of Jeffreys' scale is not a ``one size fits all" reliable tool for model comparison and that it may lead to biased conclusions. Furthermore, we discuss the importance of choosing the right basis in the context of models that are linear with respect to their parameters and how that basis affects the parameter estimation and the derived constraints.
\end{abstract}

\maketitle

\section{Introduction}
Model comparison is at the forefront of modern science, especially in an age of huge datasets and several competing theories. Cosmology has entered an era where large amounts of data will be flowing in from CMB and LSS experiments like Planck~\cite{Planck}, BOSS~\cite{BOSS}, DES~\cite{DES}, COrE~\cite{CORE}, Euclid~\cite{Euclid}, etc. This clearly raises a fundamental question: Given some cosmological observations in the form of data and some cosmological models that may depend on one or more variables, how does one choose the best model? The reason for asking this is quite obvious. Perhaps the models correspond to predictions of different and competing fundamental theories that may explain a range of phenomena. One such example is the plethora of different Dark Energy and Modified Gravity models (see e.g. Ref.~\cite{Tsujikawa:2010sc} for details) that fit the current cosmological observations more or less equally well with General Relativity, at least within the range of a few sigmas.

A common way to answer this question has been by using Bayesian inference, see Refs.~\cite{Jeffreys,Jaynes,Mackay,D'Agostini}. For instance, the usual method of comparing minimum $\chi^2$ per effective degree of freedom is not very decisive. Other methods to decide which model gives the best description, given the data, include various Information Criteria, e.g.  Akaike~\cite{Akaike} and Bayesian~\cite{Schwarz}, which are more or less justified, see~\cite{Liddle:2004nh}, and normally do not compare well among each other.

On the other hand, the Bayesian evidence is based on Bayes theorem, see Refs. \cite{Trotta:2008qt},\cite{Liddle:2009xe} for in-depth reviews, which expresses the posterior distribution ${\cal P}(u,{\cal M}|{\bf D})$ for the parameters $u$ of the model ${\cal M}$ given the data ${\bf D}$, in terms of the likelihood distribution function ${\cal L}({\bf D}| u,{\cal M})$ within a given set of priors $\pi(u,{\cal M})$
\begin{equation}\label{BayesThm}
{\cal P}(u,{\cal M}|{\bf D}) = {{\cal L}({\bf D}| u,{\cal M}) \,\pi( u,{\cal M})\over E({\bf D}|{\cal M})}\,,
\end{equation}
where the likelihood can be obtained from ${\cal L}({\bf D}| u,{\cal M})=\exp(-\chi^2(u)/2)$. We should stress that in what follows we will assume a Gaussian likelihood and, as we discuss in great detail in the next section, we will consider its unnormalized version. Here $E$ is the Bayesian evidence, i.e. the average likelihood over the priors,
\begin{equation}\label{BE}
E({\bf D}|{\cal M}) = \int d u\ {\cal L}({\bf D}| u,{\cal M})\, \pi( u,{\cal M})\,,
\end{equation}
or roughly, the probability of the data being true given the model, integrated over the whole parameter range $u$ as defined by the priors. The comparison of the models proceeds as the ratio of this quantity evaluated for the different models
\begin{equation}\label{BF}
B_{ij} \equiv {E({\bf D}|{\cal M}_i)\over E({\bf D}|{\cal M}_j)}\,,
\end{equation}
where we have assumed equal prior probabilities for the two models. This expression may naively be considered to provide a mathematical representation of Occam's razor, because more complex models tend to be less predictive, lowering their average likelihood (within the priors) in comparison with simpler, more predictive models. Complex models can only be favored if they are able to provide a significantly improved fit to the data. The Bayes factor (\ref{BF}) is then used to give evidence for (i.e.~favor) the model ${\cal M}_i$ against the model ${\cal M}_j$ using the so-called Jeffreys' scale\footnote{More details on the Jeffreys' scale can be found in later sections and specific threshold values in Table \ref{table1}.}, a particular interpretation of the Bayes factor which strengthens its verdict roughly each time the logarithm $\ln B_{ij}$ increases by one unit, from 0 (indecisive) to greater than 5 (strongly indecisive). Finally, it should be stressed that the Bayes factor has an intrinsic definition as the posterior odds ratio of the two models independent of the Jeffreys'scale \cite{kasraft}.

In Section \ref{fitting}, we explicitly calculate the Bayes factors for all models that are linear with respect to their parameters. In Section \ref{modelcomp} we discuss the case of nested models, e.g. one with $M_1$ and another with $M_2$ parameters and we derive analytic expressions for the Bayes factor while in Section \ref{secfom} we discuss the same problem for the Figure of Merit. With all this machinery and the use of the explicit example we demonstrate in Section \ref{modelcomp} that the Jeffreys' scale is not a ``one size fits  all" reliable tool for model comparison, contrary to the common belief by many people.

\section{Bayesian model comparison \label{modelcomp}}
\subsection{The Jeffreys' scale revisited}
In this section we will present results related to the Bayes factor $B_{ij}$ (see Refs \cite{Berger1}, \cite{Berger2}, \cite{Berger3}, \cite{Protassov} and references there-in for more details) in the context of our simple model. In this case, the Bayes factor $B_{ij}$ can be written as
\be
B_{ij} \equiv \frac{L(M_i)}{L(M_j)}
\ee
where $L(M_i)$ denotes the probability $p(D|M_i)$, called likelihood for the model $M_i$, to
obtain the data $D$ if the model $M_i$ is the true one. Generally, $L(M_i)$ is defined as:
\be
L(M_i)\equiv p(D|M_i) =\int da \cdot
p(a| M_i) {\cal L}_i(a)\label{evidence}
\ee
for models with one free parameter and
where $p(a| M_i)$ is the prior probability for the parameter $a$.
Also, ${\cal L}_i(a)$ is the likelihood for the parameter $a$ in
the model and \cite{John:2002gg}
\be
{\cal L}_i(a)\equiv e^{-\chi^2(a)/2}
\ee
However, at his point we should note that it is not uncommon in the community to use instead the normalized likelihood of Eq.~(\ref{likelihood1}). In this case, the only difference is just a multiplicative constant $\mathcal{N}$ given by Eq.~(\ref{norm}), which will propagate in all the subsequent calculations. We have explicitly checked that if we include the normalization our conclusions do not change.

In the case that $a$ has flat prior probabilities, that is we have
no prior information on $a$ besides that it lies in some range
$[a,a+\Delta a]$ then $p(a|M_i)=\frac{1}{\Delta a}$ and
\be
L(M_i)=\frac{1}{\Delta a}\int_a^{a+\Delta a} da e^{-\chi^2(a)/2}
\ee

Of course, all this can be easily generalized for models having more
than one parameter as follows
\be
L(M_i)=\left(\prod_{j=0}^{M-1}\frac{1}{\Delta a_j}\right)\int_{\vec{a}}^{\vec{a}+\Delta \vec{a}} e^{-\chi^2(\vec{a})/2}\cdot d\vec{a} \label{bayes1}
\ee
where $M$ is the total number of parameters and the integration over $d\vec{a}\equiv \prod_{j=0}^{M-1} d a_j=da_0 da_1 ... da_{M-1}$ is assumed to be multidimensional in general. Also, we will consider Gaussian priors of the form:
\be
Pr(\vec{a})=\frac{|H_{ij}|^{1/2}}{(2\pi)^{M/2}}~e^{- (a-a_{\rm pr})_i~H_{ij}~(a-a_{\rm pr})_j/2}, \label{GausPr}
\ee
where the $M$ priors are centered around the values $a_{\rm{pr},i}$ and $H_{ij}$ is their inverse covariance. Also, we have properly normalized the gaussian priors to unity, such that $\int_{-\infty}^{+\infty}Pr(\vec{a})d\vec{a}=1$.

The interpretation of the Bayes factor $B_{ij}$ is
that \cite{John:2002gg} when $1<B_{ij}<3$ there is evidence against
$M_j$ when compared with $M_i$, but it is only worth a bare
mention. When $3<B_{ij}<20$ the evidence against $M_j$ is definite
but not strong. For $20<B_{ij}<150$ the evidence is strong and for
$B_{ij}>150$ it is very strong. For handy reference we include the values of both the linear and the logarithmic Jeffreys' scale in Table \ref{table1}. Jeffreys in his seminal paper \cite{Jeffreys, Robertetal} provides somewhat different but in general consistent values. Several examples of the use of the Jeffreys' scale in cosmology and astronomy can by found in \cite{John:2002gg,Lazkoz:2005sp,Efstathiou:2008ed,Jenkins:2011va} and references there-in.

\subsection{The Bayesian evidence}
\subsubsection{Gaussian priors}
Using the machinery of the previous sections we will now calculate the Bayesian evidence of Eq.~(\ref{bayes1}) in the case of the Gaussian priors. Using Eqs.~(\ref{evidence}) and (\ref{GausPr}) we have:
\ba
B_1 &=&  \int_{-\infty}^{+\infty} e^{-\chi^2(\vec{a})/2} Pr(\vec{a}) \cdot  d\vec{a}\nn \\ &=& \frac{|H_{ij}|^{1/2}}{(2\pi)^{M/2}} e^{-\chi^2_{\rm min}/2}\int_{-\infty}^{+\infty} e^{- (a-a_{\rm min})_i~F_{ij}~(a-a_{\rm min})_j/2- (a-a_{\rm pr})_i~H_{ij}~(a-a_{\rm pr})_j/2}\cdot d\vec{a} \label{gaussres0}
\ea
At this point we can introduce a new matrix $G_{ij}$ and constants $c$ and $a_{1,i}$ such that:
\ba
G_{ij}&=&F_{ij}+H_{ij}  \\
a_{1,i} &=& \left(a_{k,\rm min}F_{kj}+a_{k,\rm pr}H_{kj}\right)G_{ij}^{-1}  \\
c &=&(a_{\rm min}-a_1)_i~F_{ij}~(a_{\rm min}-a_1)_j+(a_{\rm pr}-a_1)_i~H_{ij}~(a_{\rm pr}-a_1)_j\\
(a-a_1)_i~G_{ij}~(a-a_1)_j+c&=&(a-a_{\rm min})_i~F_{ij}~(a-a_{\rm min})_j+(a-a_{\rm pr})_i~H_{ij}~(a-a_{\rm pr})_j \label{newmatG}
\ea
Obviously, when the best-fit and the priors are centered in the same point, ie $a_{i,\rm min}=a_{i,\rm pr}$, then we have that $a_{1,i}=a_{i,\rm min}=a_{i,\rm pr}$ and $c=0$. By using Eqs.~(\ref{newmatG}) we can proceed with the calculation of (\ref{gaussres0}) as usual:
\ba
B_1 &=& \frac{|H_{ij}|^{1/2}}{(2\pi)^{M/2}} e^{-\chi^2_{\rm min}/2}\int_{-\infty}^{+\infty} e^{- (a-a_1)_i~G_{ij}~(a-a_1)_j/2-c/2}\cdot d\vec{a}\nn \\ &=& \frac{|H_{ij}|^{1/2}}{(2\pi)^{M/2}} e^{-\chi^2_{\rm min}/2-c/2} \frac{(2\pi)^{M/2}}{|G_{ij}|^{1/2}} \nn \\ &=& e^{-\chi^2_{\rm min}/2-c/2} \frac{|H_{ij}|^{1/2}}{|G_{ij}|^{1/2}}\nn \\ &=& e^{-\chi^2_{\rm min}/2-c/2} |I_M+H^{-1} F|^{-1/2} \label{gaussres1}
\ea where in the second line we used Eqs.~(\ref{transf1})-(\ref{transf3}), in the last line we used the well known matrix identity $|X+A|=|X||I_M+X^{-1}A|$ where $I_M$ is the $M \times M$ unit matrix, and finally we used the fact that $G_{ij}=H_{ij}+F_{ij}$. By using the matrix identity $|A|=\frac{1}{2}\left(tr(A)^2-tr(A^2)\right)$, where $tr(A)$ is the trace of the matrix $A$, we can express the determinant of a sum of the unit matrix $I_M$ and a matrix $B$ as
\ba
|I_M+B|&=& \frac{1}{2}\left(\tr(I_M+B)^2-\tr((I_M+B)\cdot(I_M+B))\right) \nn \\
&=& \frac{M(M-1)}{2} +(M-1) \tr(B)+|B|
\ea
and then, the expression $|I_M+H^{-1} F|$ can be expanded to
\be
|I_M+H^{-1} F|=\frac{M(M-1)}{2} +(M-1) \tr(H^{-1} F)+|H^{-1}||F|
\ee

Then the Bayes factor can be written as
\ba
B_{12}&=& e^{-\Delta\chi^2_{\rm 1,2 min}/2-\Delta c_{1,2}/2}\frac{|I_{M_1}+H_{(1)}^{-1} F_{(1)}|^{-1/2}}{|I_{M_2}+H_{(2)}^{-1} F_{(2)}|^{-1/2}} \nn \\
&=&e^{-\Delta\chi^2_{\rm 1,2 min}/2-\Delta c_{1,2}/2}\left(\frac{\frac{M_1(M_1-1)}{2} +(M_1-1) \tr(H_{(1)}^{-1} F_{(1)})+|H_{(1)}^{-1}||F_{(1)}|}{\frac{M_2(M_2-1)}{2} +(M_2-1) \tr(H_{(2)}^{-1} F_{(2)})+|H_{(2)}^{-1}||F_{(2)}|}\right)^{-1/2}
\label{bayesgaussresult}\ea
where $\Delta\chi^2_{\rm 1,2 min}=\chi^2_{\rm 1 min}-\chi^2_{\rm 2 min}$ and $\Delta c_{1,2}=c_1-c_2$ are the values of the constant $c$ for the two models. Also, in the last line we have labeled all the different quantities with $(1)$ and $(2)$ to indicate the two models $1$ and $2$ respectively. Finally, it should be noted that Eq.~(\ref{bayesgaussresult}) is an \textit{exact} result.

\begin{table}
\begin{center}
\caption{The values of both the linear and the logarithmic Jeffreys' scale, and the AIC and BIC criteria. For references on these values check the text.} \label{table1}
\begin{tabular}{cccc}
  \hline
  \hline
  \hspace{5pt} $B_{ij}$ \hspace{5pt} & \hspace{5pt} $\ln{B_{ij}}$ \hspace{5pt}&\hspace{5pt} Evidence \hspace{5pt}\\
  \hline
  \hspace{5pt}$1\leq B_{ij}<3$  \hspace{5pt} &\hspace{5pt} $0\leq B_{ij}<1.1$ \hspace{5pt}&\hspace{5pt} Weak       \\
  \hspace{5pt}$3\leq B_{ij}<20$ \hspace{5pt} &\hspace{5pt} $1.1\leq B_{ij}<3$   \hspace{5pt}&\hspace{5pt} Definite   \\
  \hspace{5pt}$20\leq B_{ij}<150$\hspace{5pt} &\hspace{5pt} $3\leq B_{ij}<5$   \hspace{5pt}&\hspace{5pt} Strong     \\
  \hspace{5pt}$150\leq B_{ij}$\hspace{5pt} &\hspace{5pt} $5\leq B_{ij}$   \hspace{5pt}&\hspace{5pt} Very Strong\\
  \hline
\end{tabular}
\end{center}
\end{table}

\subsubsection{Flat priors}

Alternatively, we can choose our priors to be top-hat and centered around the best-fit, ie we integrate in the range $[\vec{a}_{\rm min}-\frac{\Delta \vec{a}}{2},\vec{a}_{\rm min}+\frac{\Delta \vec{a}}{2}]$, with our flat top-hat priors being equal to $\Delta \vec{a}=\overrightarrow{const.}\in R^M$, where $M$ is as usual the total number of parameters of the model\footnote{ However, it should be noted that especially in a large number of dimensions, flat priors are well known to give misleading results.}.

In this case then, by using Eq.~(\ref{bayes1}), the evidence $B_1$ for the model can be written as
\ba
B_1 &\!=\!&  \left(\prod_{j=0}^{M-1}\frac{1}{\Delta a_j}\right)\int_{\vec{a}_{\rm min}-\Delta \vec{a}/2}^{\vec{a}_{\rm min}+\Delta \vec{a}/2} e^{-\chi^2(\vec{a})/2}\cdot d\vec{a}\nn \\
&\!=\!&  \left(\prod_{j=0}^{M-1}\frac{1}{\Delta a_j}\right)\,e^{-\chi^2_{\rm min}/2}\int_{\vec{a}_{\rm min}-\Delta \vec{a}/2}^{\vec{a}_{\rm min}+\Delta \vec{a}/2} e^{- (a-a_{\rm min})_i~F_{ij}~(a-a_{\rm min})_j/2}\cdot d\vec{a} \label{B1def}
\ea
Using Eqs.~(\ref{chi2def1}) and (\ref{transf1})-(\ref{transf3}) we can define a transverse of the Fisher matrix, $F^\perp \equiv U^{-1}F\,U$, where $U$ is the unitary off-diagonal $M$-dimensional matrix, e.g.  for $M=2$, $$U=\left(\begin{array}{cc} 0 & 1\\ 1 & 0\end{array}\right),$$ as well as the transverse of the Cholesky decomposition of $F^\perp$, denoted with the matrix $D^\perp$. With this, we can write (\ref{B1def}) as
\ba
B_1&\!\approx\!&  \left(\prod_{j=0}^{M-1}\frac{1}{\Delta a_j}\right)e^{-\chi^2_{\rm min}/2} \left|F\right|^{-1/2} \left(2\pi\right)^{M/2} \prod_{i=0}^{M-1}\erf\left(\frac{D^\perp_{ii}\Delta a_i}{2\sqrt{2}}\right)\nn \\ &=&
\frac{1}{\mathcal{N}_1 V_{1,prior}}\prod_{i=0}^{M_1-1}\erf\left(\frac{D_{ii}^{\perp(1)}\Delta a_i^{(1)} }{2\sqrt{2} }\right)\,,\label{result0}
\ea
where we have used Eq.~(\ref{norm}) and we have defined $V_{1,prior}\equiv \prod_{j=0}^{M_1-1}\Delta a_j^{(1)}$ as the ``volume" of our priors for this model 1. The function $\erf(x)$ is the usual error function defined as $\erf(x)\equiv \frac{2}{\sqrt{\pi}}\int_0^x e^{-t^2}dt$, see Ref. \cite{handbook} for more details. Then the Bayes factor for two models based on Eq.~(\ref{model}), labeled $1$ and $2$ with a total number of parameters $M_1$ and $M_2$, is
\be
B_{12}= \frac{\mathcal{N}_2 V_{2,prior}}{\mathcal{N}_1 V_{1,prior}}\cdot \frac{\prod_{i=0}^{M_1-1}\erf\left(\frac{D_{ii}^{\perp(1)}\Delta a_i^{(1)} }{2\sqrt{2}}\right)}{\prod_{i=0}^{M_2-1}\erf\left(\frac{D_{ii}^{\perp(2)}\Delta a_i^{(2)} }{2\sqrt{2}}\right)}\,. \label{result1}
\ee

Also, in the last line we have labeled all the different quantities with a $(...)^{(1)}$ to indicate model $1$. At this point we will now consider two different cases:
\begin{itemize}
\item When the priors are much smaller than the errors of the best-fit parameters, ie the arguments of the error functions are small, or $D_{ii}^{\perp}\Delta a_i\ll1$.
\item When the priors are much larger than the errors of the best-fit parameters, ie the arguments of the error functions are large, or  $D_{ii}^{\perp}\Delta a_i\gg1$.
\end{itemize}
The following expansions of the error function are useful:
\ba \erf(x)& \approx & \frac{2\sqrt{2}~ x}{\sqrt{2\pi}}\left(1-\frac{x^2}{3}\right)+... ~~~ \textrm{for}~~~x\ll1 \nn \\ \erf(x)& \approx & 1-\frac{e^{-x^2}}{\sqrt{\pi } x}+... ~~~ \textrm{for}~~~x\gg1 \ea
In the first case (when $x\ll1$), the evidence $B_1$ of Eq.~(\ref{result0}) becomes
\ba
B_1&\approx& e^{-\chi^2_{\rm min}/2}\prod_{j}^{M-1}\left(1-\frac{(D^\perp_{jj}\Delta a_j)^2}{4!}+...\right)  \nn \\ &=& e^{-\chi^2_{\rm min}/2}\left(1-\sum_{j}^{M-1}\frac{(D^\perp_{jj}\Delta a_j)^2}{4!}+...\right)
\ea
where we have used the fact that $\prod_{j=0}^{M-1}D^\perp_{jj}=|D^\perp|=|F|^{1/2}$. Then the Bayes factor can be written as
\ba
B_{12}&\approx& e^{-\Delta\chi^2_{\rm 1,2 min}/2}\frac{\prod_{j=0}^{M_1-1}\left(1-\frac{(D^{\perp(1)}_{jj}\Delta a_j^{(1)})^2}{4!}+...\right)}{\prod_{j=0}^{M_2-1}\left(1-\frac{(D^{\perp(2)}_{jj}\Delta a_j^{(2)})^2}{4!}+...\right)} \nn \\ &=& e^{-\Delta\chi^2_{\rm 1,2 min}/2}\left(1-\sum_{j=0}^{M_1-1}\frac{(D^{\perp(1)}_{jj}\Delta a_j^{(1)})^2}{4!}+ \sum_{j=0}^{M_2-1}\frac{(D^{\perp(2)}_{jj}\Delta a_j^{(2)})^2}{4!} +...\right) \label{result2}
\ea
where $\Delta\chi^2_{\rm 1,2 min}=\chi^2_{\rm 1 min}-\chi^2_{\rm 2 min}$. Also, in the last line we have labeled all the different quantities with $(...)^{(1)}$ and $(...)^{(2)}$ to indicate the models $1$ and $2$ respectively. In this limit the second term of the expression is expected to be close to $1$.

In the second case (when $x\gg1$) the evidence becomes \be B_1\approx \left(\prod_{j=0}^{M-1}\frac{1}{\Delta a_j}\right)e^{-\chi^2_{\rm min}/2} \left|F\right|^{-1/2} \left(2\pi\right)^{M/2} \left(1-\frac{1}{(2\pi)^{1/2}}\sum_{j=0}^{M-1}\frac{e^{-(D^\perp_{jj}\Delta a_j)^2/8}}{D^\perp_{jj}\Delta a_j/4}\right)\ee where in this limit the last term of the expression is expected to be close to $1$. Then the Bayes factor is just $B_{12}=\frac{B_1}{B_2}$.

\begin{figure*}[t!]
\centering
\vspace{0cm}\rotatebox{0}{\vspace{0cm}\hspace{0cm}\resizebox{0.95\textwidth}{!}{\includegraphics{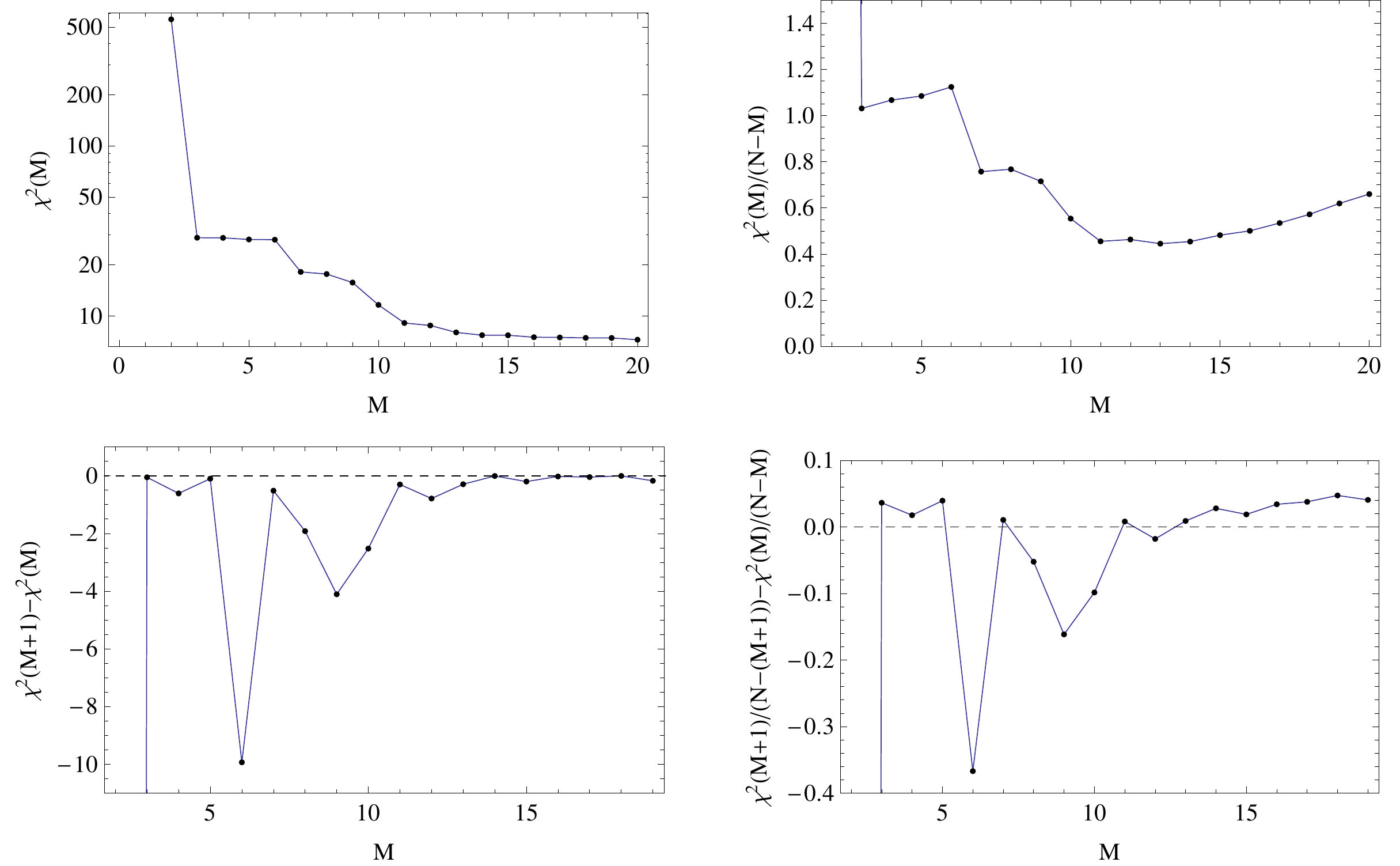}}}
\caption{The best-fit $\chi^2$ as a function of the number of parameters $M$ (top left), the best-fit $\chi^2$ per degree of freedom $N-M$ as a function of the number of parameters $M$ (top right), the difference in the best-fit $\chi^2$ between two models with parameters $M_1=M+1$ and $M_2=M$ (bottom left) and the the difference in the best-fit $\chi^2$ per degree of freedom between two models with parameters $M_1=M+1$ and $M_2=M$ (bottom right). \label{plots1}}
\end{figure*}

\subsection{Analysis\label{analysis}}
As it can be seen from Eqs.~(\ref{bayesgaussresult}) and (\ref{result1}), in both cases (gaussian and flat priors) the Bayes factor for this class of models can be written as
\be
B_{12}=e^{-\Delta\chi^2_{\rm 1,2 min}/2}\cdot G(M_1,M_2)
\ee
where the function $G(M_1,M_2)$ contains all the extra information of the models via their covariance matrices. Then the logarithmic Bayes factor can be written as
\be
\ln(B_{12})=-\Delta\chi^2_{\rm 1,2 min}/2 +\ln(G(M_1,M_2)) \label{Bayesfac}
\ee
For example,  in this order of the approximation and in the case of the flat priors Eq.~(\ref{result2}) gives
\be
\ln(B_{12})=-\Delta\chi^2_{\rm 1,2 min}/2 -\sum_{j=0}^{M_1-1}\frac{(D^{\perp(1)}_{jj}\Delta a_j^{(1)})^2}{4!}+ \sum_{j=0}^{M_2-1}\frac{(D^{\perp(2)}_{jj}\Delta a_j^{(2)})^2}{4!} +...
\label{reslnbij}\ee
As it can be seen, Eq.~(\ref{reslnbij}) does not only contain the difference between the $\chi^2_{min}$ of the two models but also contains information on their covariances via the the last two terms. Clearly, these two terms \textit{depend strongly} on the data and the model at hand, thus introducing a further complexity in the model comparison.

To back up our claims we also present an explicit example. First, we created a set of 31 mock data points $(x_i, y_i, \sigma_{y_i})$ with noise in the range $x\in[0.025,1.55]$ based on the same model as in \cite{Nesseris:2012tt}
\be
f(x)=a + (x - b) \exp(-c~x^2),\label{exmodel}
\ee
where the parameters $(a,b,c)$ have the values $(0.25,0.25,0.5)$ respectively. The choice of the model was completely ad hoc, except for the requirement to be well behaved (smooth) and that it exhibits some interesting features, such as a maximum at some point $x$. Then, we \textit{fit these data with two polynomials} that have a different number of parameters $M_1$ and $M_2$, e.g. $M_2>M_1$ or $M_1>M_2$. In other words, our model comparison is done between the models $M_1$ and $M_2$ and not between $f(x)$ of Eq. (\ref{exmodel}) and a polynomial.

\begin{figure*}[t!]
\centering
\vspace{0cm}\rotatebox{0}{\vspace{0cm}\hspace{0cm}\resizebox{0.45\textwidth}{!}{\includegraphics{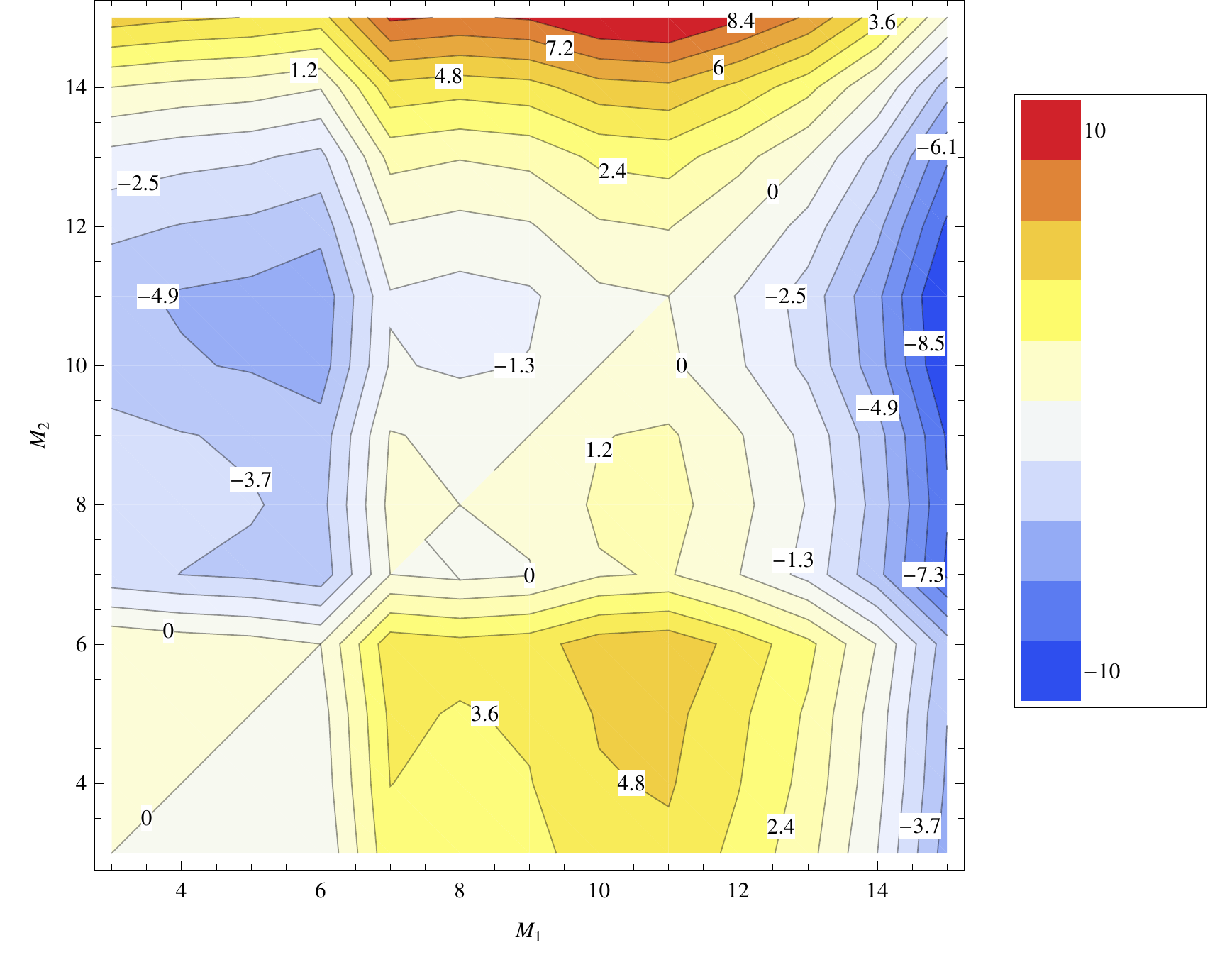}}}
\vspace{0cm}\rotatebox{0}{\vspace{0cm}\hspace{0cm}\resizebox{0.45\textwidth}{!}{\includegraphics{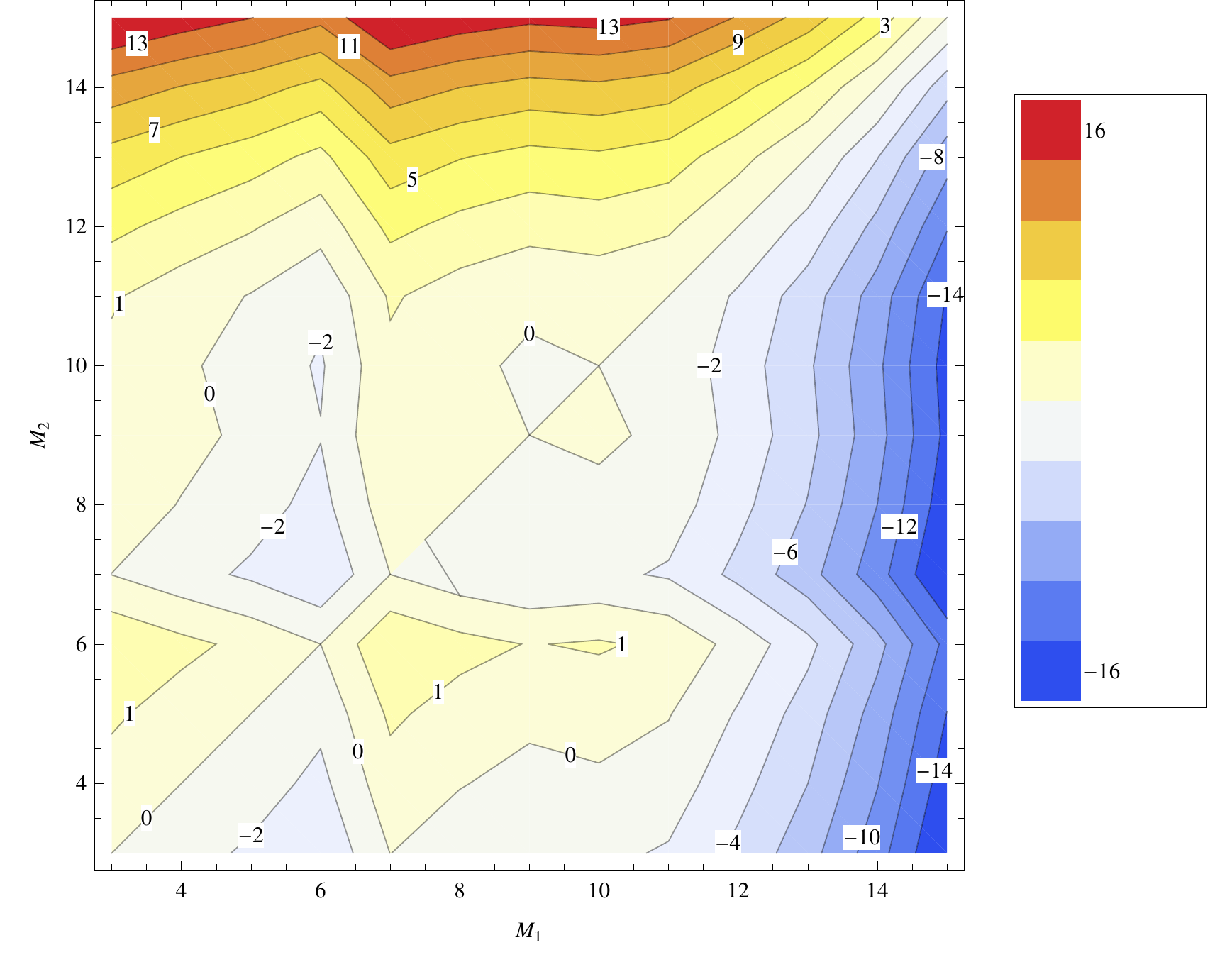}}}
\caption{Contour plots of the Bayes factor $\log{B_{12}}$ of Eq.~(\ref{Bayesfac}), calculated in the case of flat priors by using Eq.~(\ref{result1}) when the priors are taken for simplicity to be proportional to the errors of the best parameters $\Delta a_i=n~\sigma_i$ for $n=3$ (left) and $n=7$ (right). The red color corresponds to a high value for the Bayes factor $\log{B_{12}}$, ie model $M_1$ preferred, blue to a low (negative) value for the Bayes factor $\log{B_{12}}$, ie model $M_2$ preferred, while white corresponds to equal evidence for both models. The relevant values of the Jeffreys' scale are given in Table \ref{table1}. \label{plots2}}
\end{figure*}

\begin{figure*}[t!]
\centering
\vspace{0cm}\rotatebox{0}{\vspace{0cm}\hspace{0cm}\resizebox{0.45\textwidth}{!}{\includegraphics{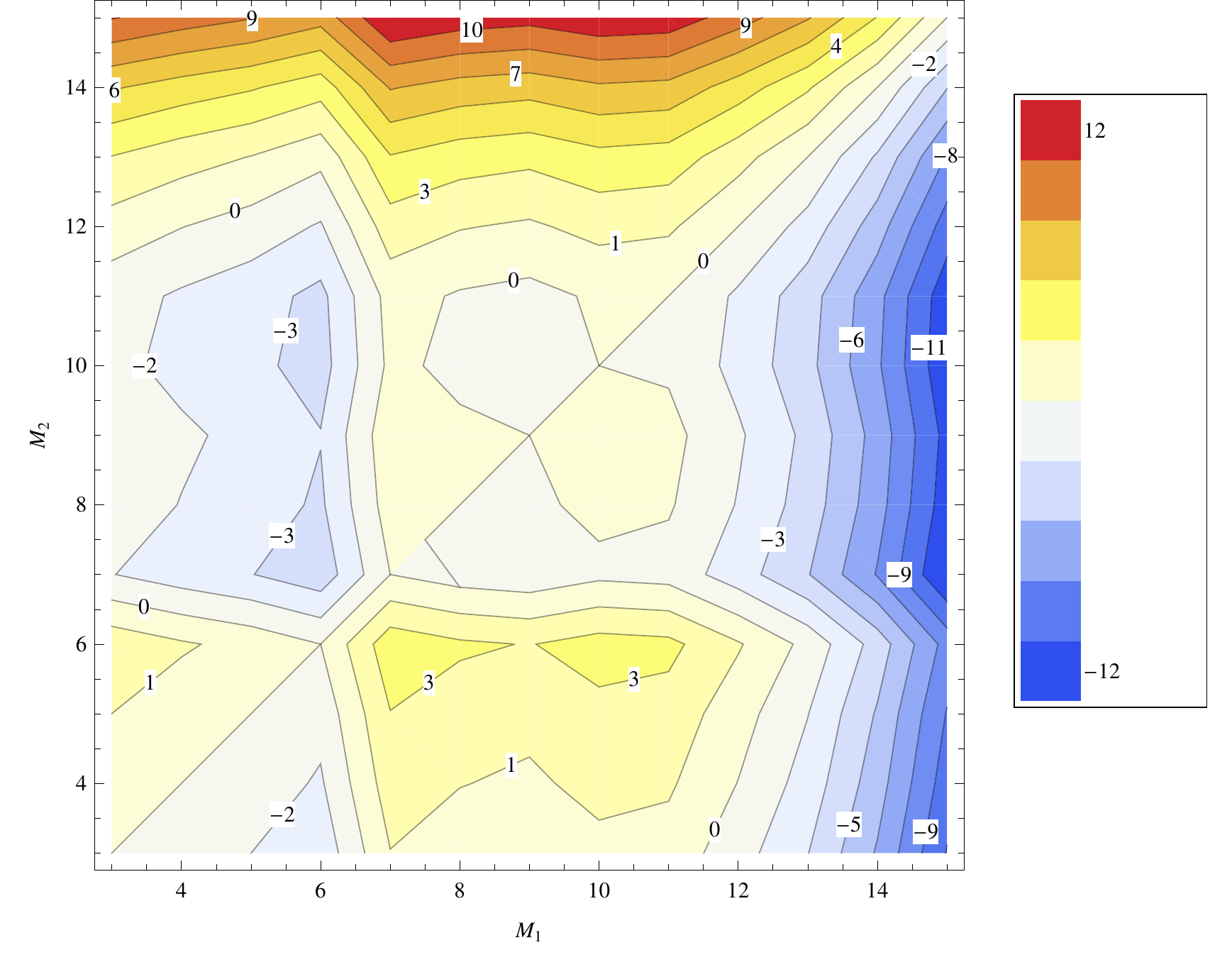}}}
\vspace{0cm}\rotatebox{0}{\vspace{0cm}\hspace{0cm}\resizebox{0.45\textwidth}{!}{\includegraphics{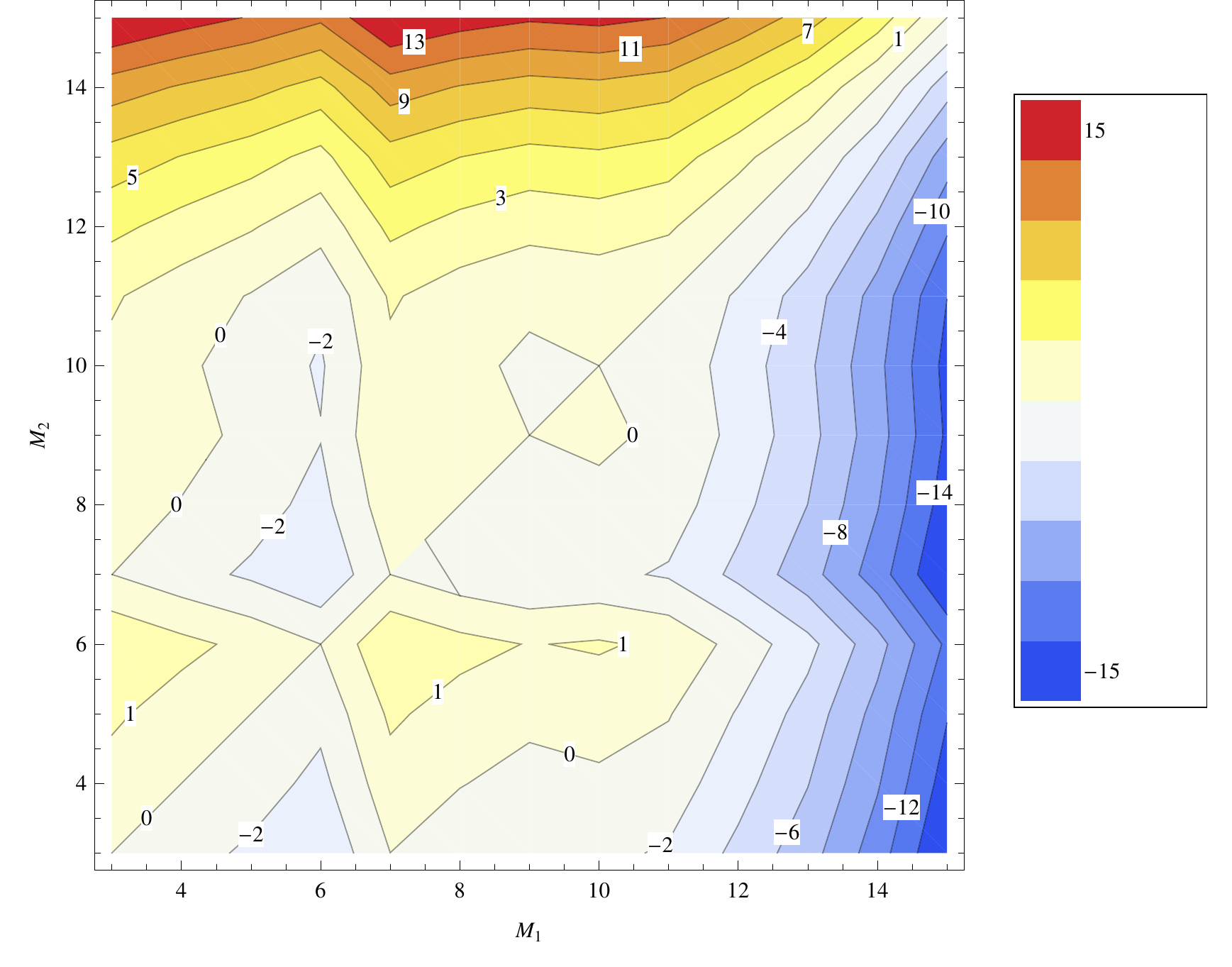}}}
\caption{Contour plots of the Bayes factor $\log{B_{12}}$ of Eq.~(\ref{Bayesfac}), calculated in the case of gaussian priors by using Eq.~(\ref{bayesgaussresult}) when the priors are taken for simplicity to be proportional to the errors of the best parameters $H_{ii}=n~\sigma_i^2$ and $H_{ij}=0$ for $i\neq j$, when $n=3$ (left) and $n=7$ (right). The red color corresponds to a high value for the Bayes factor $\log{B_{12}}$, ie model $M_1$ preferred, blue to a low (negative) value for the Bayes factor $\log{B_{12}}$, ie model $M_2$ preferred, while white corresponds to equal evidence for both models. The relevant values of the Jeffreys' scale are given in Table \ref{table1}. \label{plots3}}
\end{figure*}

In Fig.~\ref{plots1} we show the best-fit $\chi^2$ as a function of the number of parameters $M$ (top left), the best-fit $\chi^2$ per degree of freedom $N-M$ as a function of the number of parameters $M$ (top right), the difference in the best-fit $\chi^2$ between two models with parameters $M_1=M+1$ and $M_2=M$ (bottom left) and the the difference in the best-fit $\chi^2$ per degree of freedom between two models with parameters $M_1=M+1$ and $M_2=M$ (bottom right). While the absolute value of the $\chi^2$ can be decreased almost arbitrarily by increasing the parameters $M$, the improvement at some point ceases to become relevant compared to a model with with $M-1$ parameters, ie the fit does not become better with respect to a model with just one less parameter. Also, the best-fit $\chi^2$ per degree of freedom $N-M$ seems to have a minimum, in this case for $M=13$, which means that beyond that point adding more parameters just does not increase the quality of the fit.

In Fig.~\ref{plots2} we show contour plots of the Bayes factor $\log{B_{12}}$ of Eq.~(\ref{Bayesfac}), calculated in the case of flat priors by using Eq.~(\ref{result1}) when the priors are taken for consistency to be proportional to the errors of the best parameters $\Delta a_i=n~\sigma_i$ for $n=3$ (left) and $n=7$ (right). The red color corresponds to a high value for the Bayes factor $\log{B_{12}}$, ie model $M_1$ preferred, blue to a low (negative) value for the Bayes factor $\log{B_{12}}$, ie model $M_2$ preferred, while white corresponds to equal evidence for both models. The relevant values of the Jeffreys' scale are given in Table \ref{table1}.

As it can be seen, cases that one would normally expect for $M_2$ to be ruled out, eg $M_1=4$ and $M_2=14$, as the difference in parameters is a staggering $M_2-M_1=10$ thus giving $M_2$ a big disadvantage, has a Bayes factor $\log{B_{12}}=1.2$ for $n=3$ and is actually allowed by the Jeffrey's scale! Another similar example can be seen for $M_1=4$, $M_2=10$ and $n=7$, see Fig.~\ref{plots2} on the right, where the Bayes factor is $\log{B_{12}=0}$ meaning that these two models are totally equivalent!  As it can be seen in the two plots of Fig.~\ref{plots2} the results and the conclusions for the two models $M_1$ and $M_2$ are very sensitive in the choice of the priors $\Delta a_i$.

In Fig.~\ref{plots3} we show contour plots of the Bayes factor $\log{B_{12}}$ of Eq.~(\ref{Bayesfac}), calculated in the case of gaussian priors by using Eq.~(\ref{bayesgaussresult}) when the priors are taken for simplicity to be proportional to the errors of the best parameters $H_{ii}=n~\sigma_i^2$ and $H_{ij}=0$ for $i\neq j$, when $n=3$ (left) and $n=7$ (right). The red color corresponds to a high value for the Bayes factor $\log{B_{12}}$, ie model $M_1$ preferred, blue to a low (negative) value for the Bayes factor $\log{B_{12}}$, ie model $M_2$ preferred, while white corresponds to equal evidence for both models. The relevant values of the Jeffreys' scale are given in Table \ref{table1}. We find similar results as in the case of the flat priors, ie models that should be excluded by the Jeffrey's scale are in fact allowed.

\begin{figure*}[t!]
\centering
\vspace{0cm}\rotatebox{0}{\vspace{0cm}\hspace{0cm}\resizebox{0.47\textwidth}{!}{\includegraphics{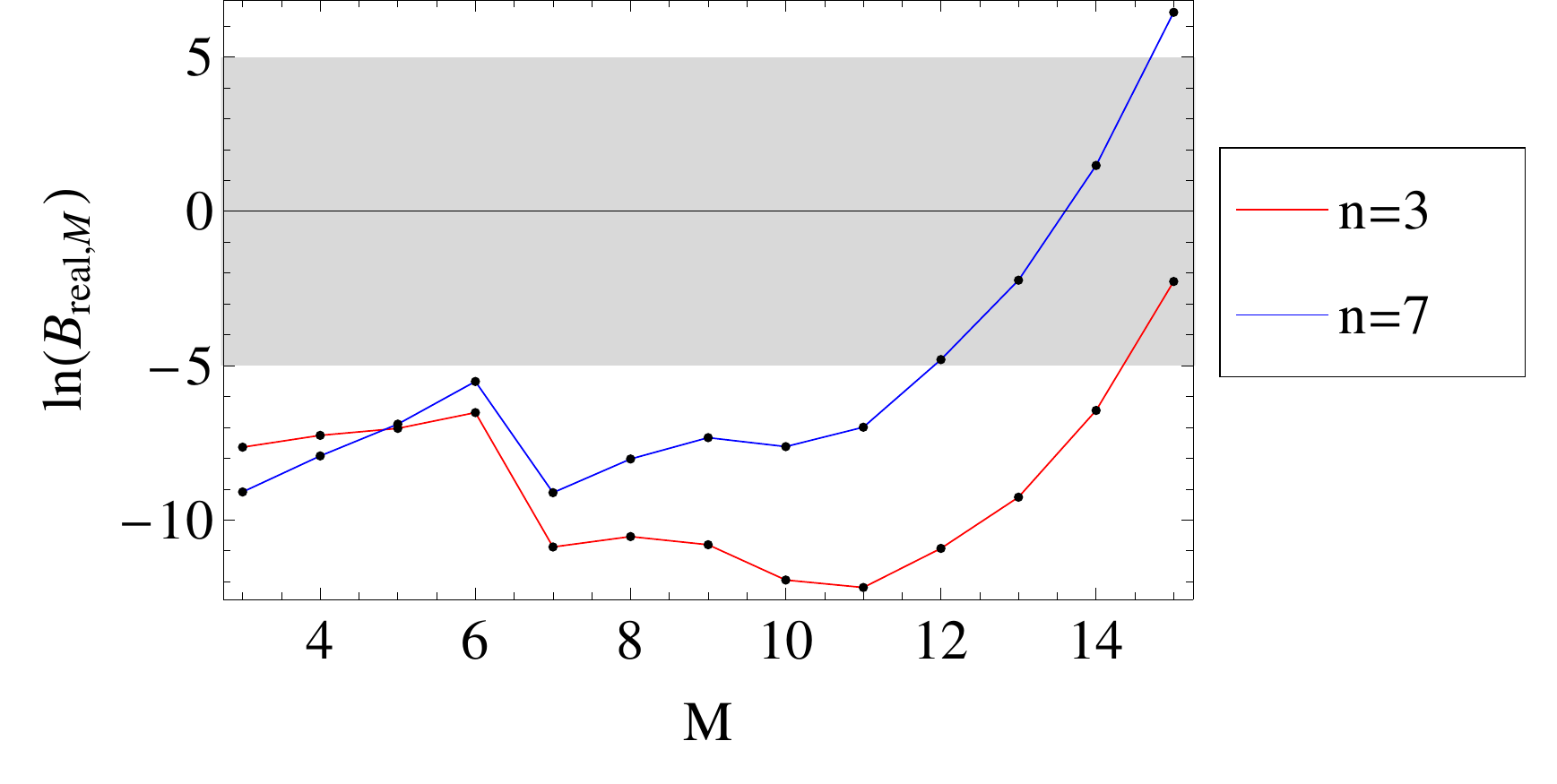}}}
\vspace{0cm}\rotatebox{0}{\vspace{0cm}\hspace{0cm}\resizebox{0.47\textwidth}{!}{\includegraphics{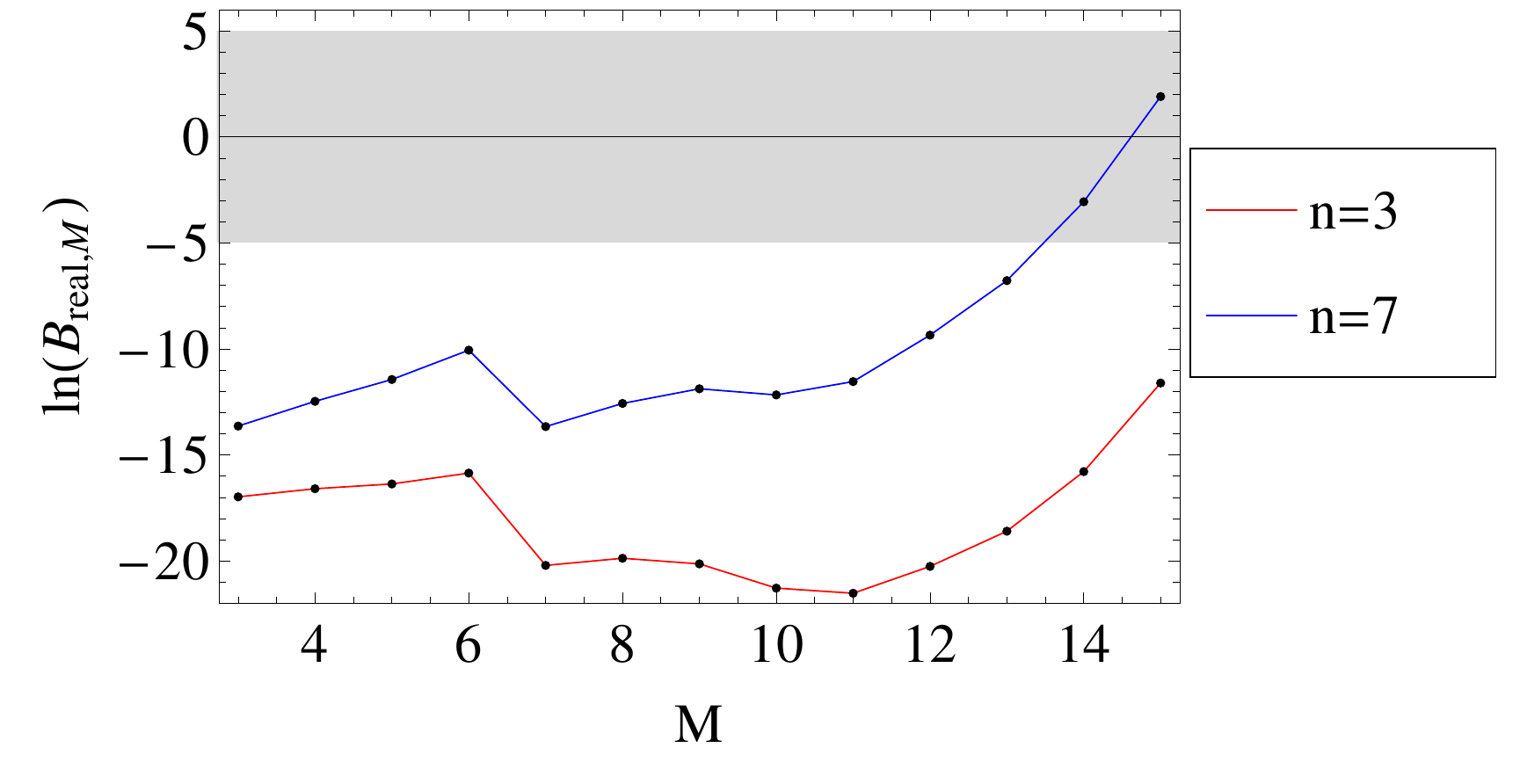}}}
\caption{The Bayes factor $\log{B_{real,M}}$ for flat (left) and gaussian (right) priors for $n=3$ and $n=7$, comparing the real model $f(x)$ of Eq. (\ref{exmodel}) vs polynomials of degree $M$. Assuming the standard thresholds of the Jeffrey's scale, there's strong support against the real model even for polynomials of very high order $M\sim 10$.\label{plotsM12vsreal}}
\end{figure*}

Also, we test the polynomial models against the "real" model of Eq.~(\ref{exmodel}). In Fig.~\ref{plotsM12vsreal} we show the Bayes factor $\log{B_{real,M}}=\log{\frac{B_{real}}{B_M}}$ for flat (left) and gaussian (right) priors for $n=3$ and $n=7$, comparing the real model $f(x)$ of Eq. (\ref{exmodel}) vs polynomials of degree $M$. Assuming the standard thresholds of the Jeffrey's scale, there is strong support against the real model even for polynomials of very high order $M\sim 10$! This clearly demonstrates that the Jeffrey's scale is an inadequate tool for model comparison, since it completely fails even in this simple example.

Finally, we have explicitly checked our methodology with other models as well and we get similar results. Specifically, we also considered the case where the ``real" model $f(x)$ is a parabola with three parameters $(a,b,c)$:
\be f(x)=a+b~x+c~x^2,\label{exmodel1}
\ee

\begin{figure*}[t!]
\centering
\vspace{0cm}\rotatebox{0}{\vspace{0cm}\hspace{0cm}\resizebox{0.45\textwidth}{!}{\includegraphics{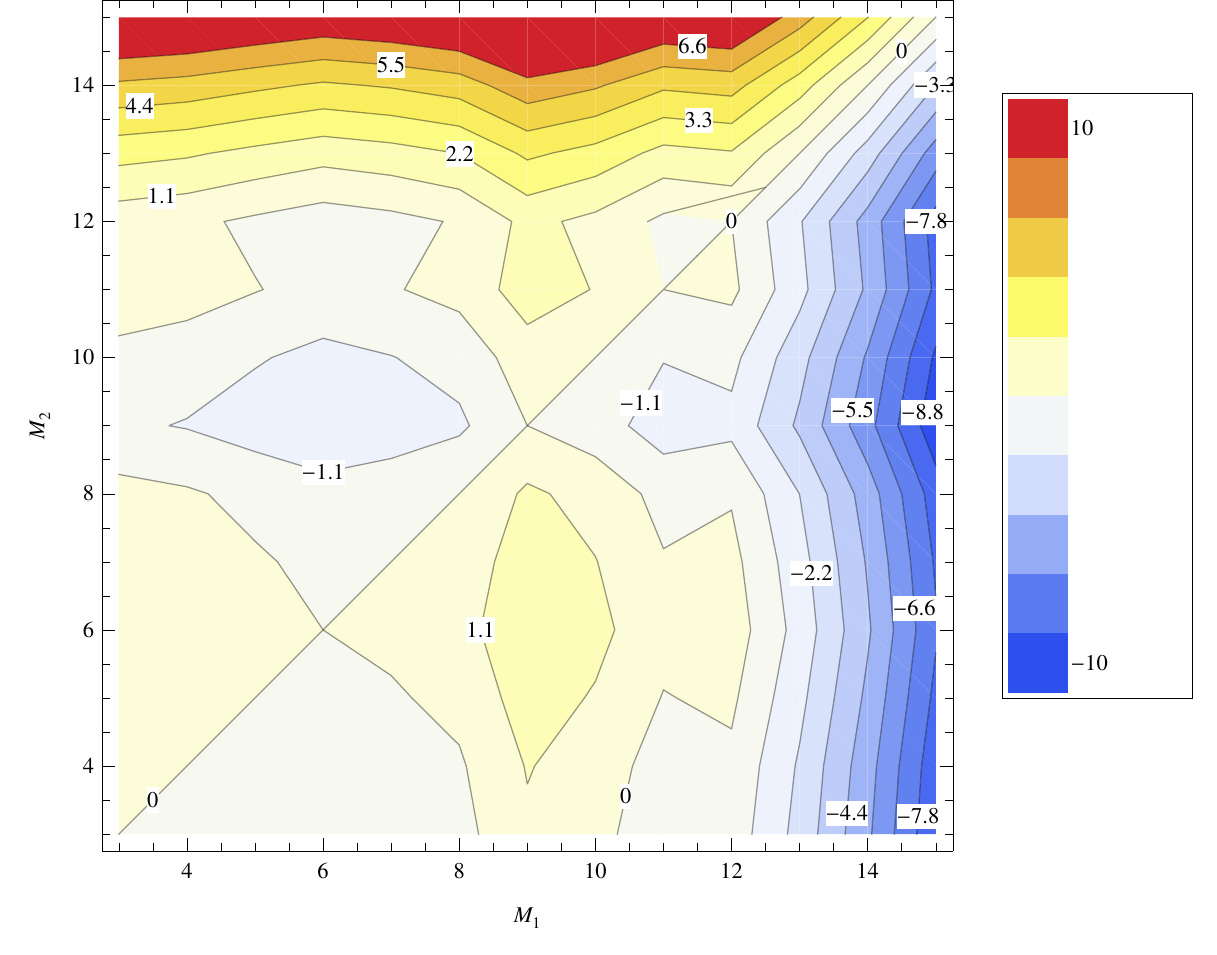}}}
\vspace{0cm}\rotatebox{0}{\vspace{0cm}\hspace{0cm}\resizebox{0.45\textwidth}{!}{\includegraphics{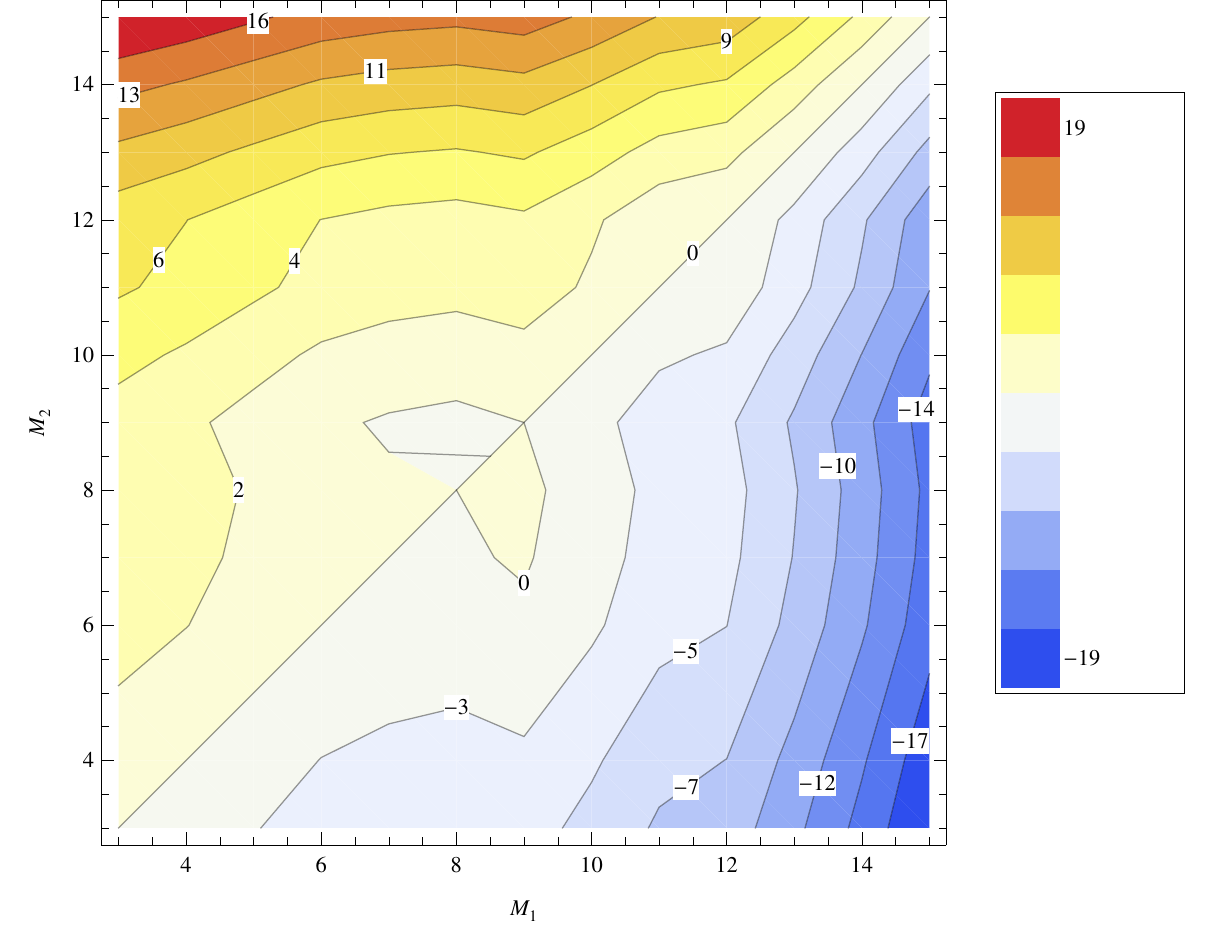}}}
\caption{Contour plots of the Bayes factor $\log{B_{12}}$ of Eq.~(\ref{Bayesfac}), in the case where the real model is given by  Eq.~(\ref{exmodel1}), calculated in the case of flat priors by using Eq.~(\ref{result1}) when the priors are taken for simplicity to be proportional to the errors of the best parameters $\Delta a_i=n~\sigma_i$ for $n=3$ (left) and $n=7$ (right). The red color corresponds to a high value for the Bayes factor $\log{B_{12}}$, ie model $M_1$ preferred, blue to a low (negative) value for the Bayes factor $\log{B_{12}}$, ie model $M_2$ preferred, while white corresponds to equal evidence for both models. The relevant values of the Jeffreys' scale are given in Table \ref{table1}. \label{plots2other}}
\end{figure*}

\begin{figure*}[t!]
\centering
\vspace{0cm}\rotatebox{0}{\vspace{0cm}\hspace{0cm}\resizebox{0.45\textwidth}{!}{\includegraphics{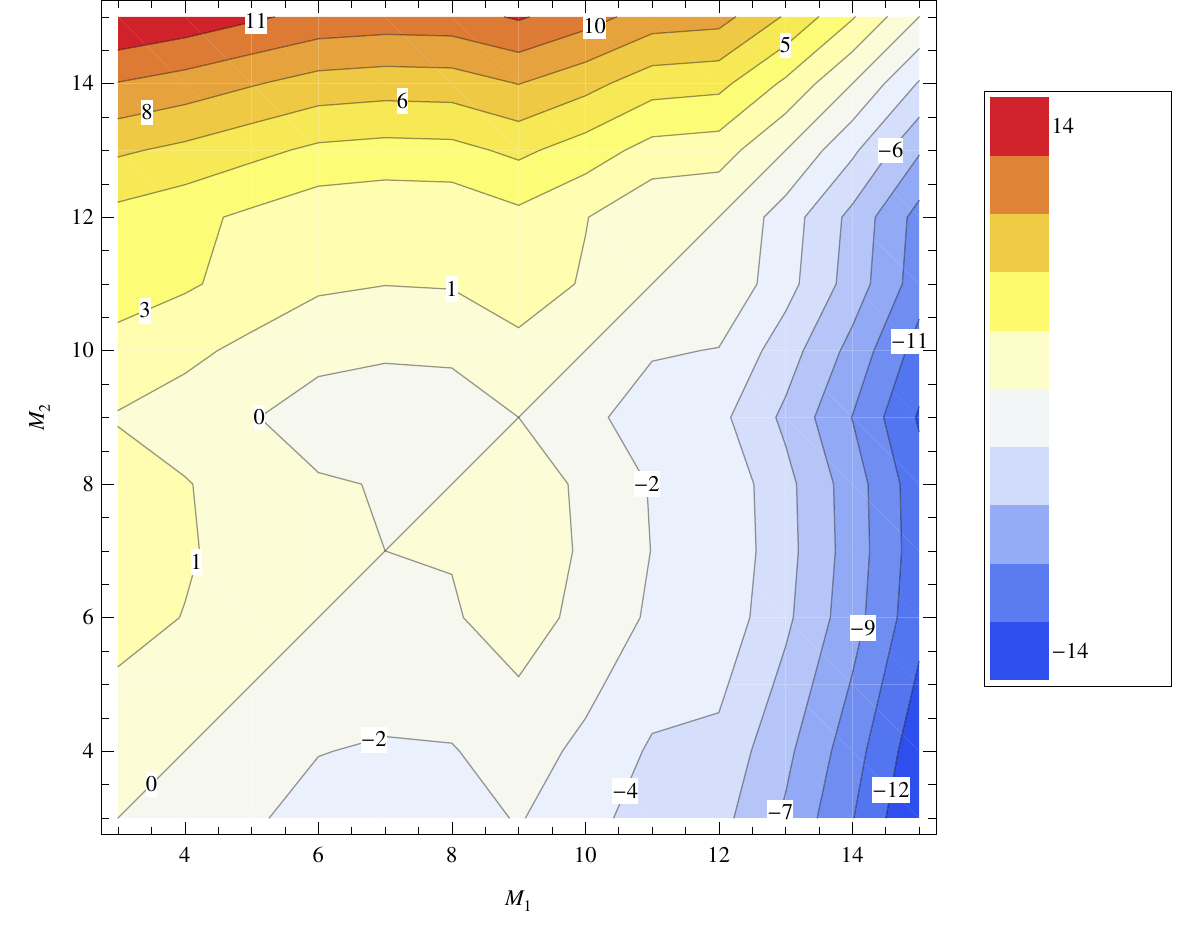}}}
\vspace{0cm}\rotatebox{0}{\vspace{0cm}\hspace{0cm}\resizebox{0.45\textwidth}{!}{\includegraphics{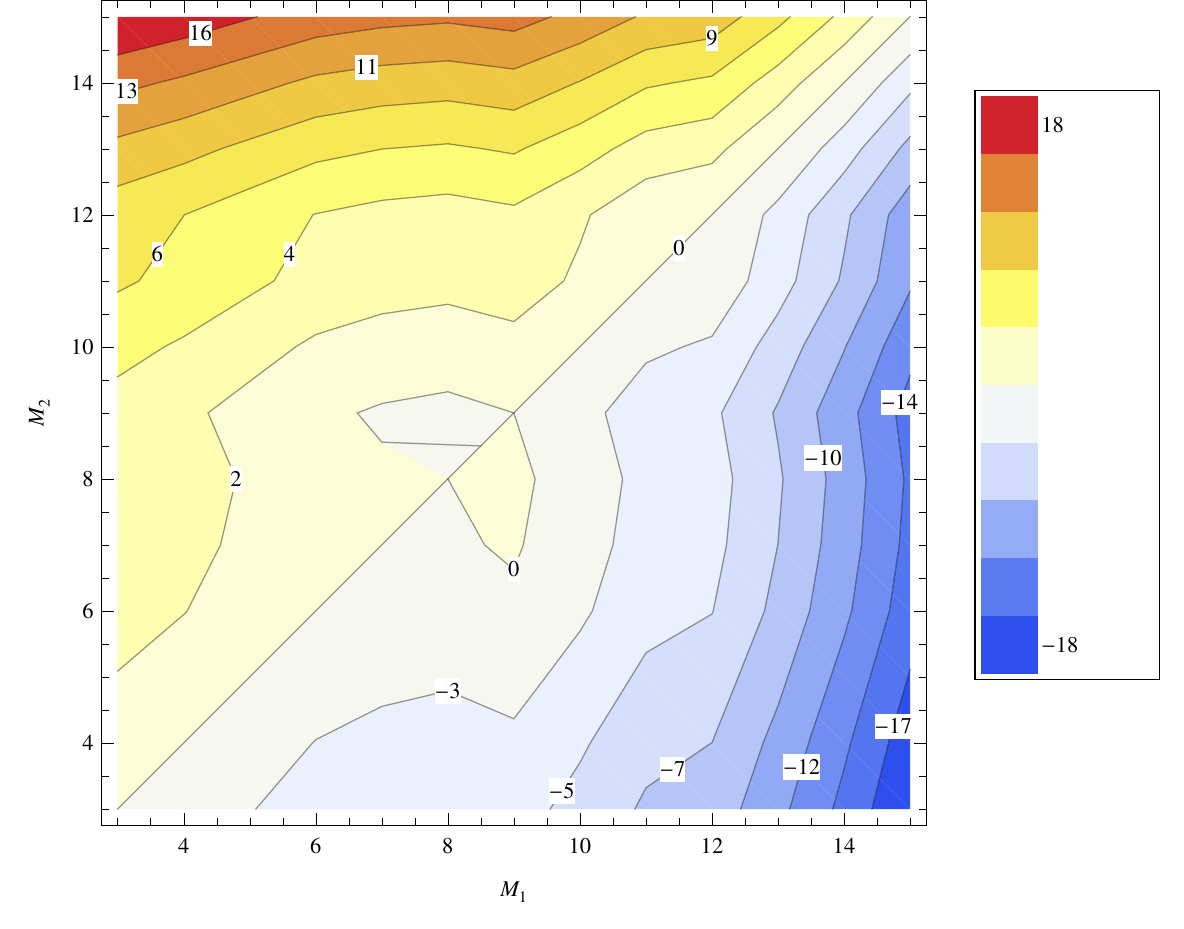}}}
\caption{Contour plots of the Bayes factor $\log{B_{12}}$ of Eq.~(\ref{Bayesfac}), in the case where the real model is given by  Eq.~(\ref{exmodel1}), calculated in the case of gaussian priors by using Eq.~(\ref{bayesgaussresult}) when the priors are taken for simplicity to be proportional to the errors of the best parameters $H_{ii}=n~\sigma_i^2$ and $H_{ij}=0$ for $i\neq j$, when $n=3$ (left) and $n=7$ (right). The red color corresponds to a high value for the Bayes factor $\log{B_{12}}$, ie model $M_1$ preferred, blue to a low (negative) value for the Bayes factor $\log{B_{12}}$, ie model $M_2$ preferred, while white corresponds to equal evidence for both models. The relevant values of the Jeffreys' scale are given in Table \ref{table1}. \label{plots3other}}
\end{figure*}

In Figs.~\ref{plots2other}, \ref{plots3other} we show contour plots of the Bayes factor $\log{B_{12}}$ of various polynomials of degrees $M_1$ and $M_2$, based on data created by using the low order polynomial of Eq.~(\ref{exmodel1}) both for flat, shown in Fig.~\ref{plots2other}, and gaussian priors, shown in Fig.~\ref{plots3other}, for n=3 (left) and n=7 (right). We again find the surprising result that cases that one would normally expect for $M_2$ to be ruled out, eg $M_1=4$ and $M_2=14$, as the difference in parameters is a staggering $M_2-M_1=10$ thus giving $M_2$ a big disadvantage, has a Bayes factor $\log{B_{12}}\sim1$ for $n=3$ that is actually allowed by the Jeffrey's scale! So, even when the real model is a function with three parameters, the Jeffrey's scale \textit{fails to rule out} a model with 14 parameters!

However, at this point we should clarify the fact that we do not criticize the Bayes factor, instead we express our disagreement with the absolute values of the Jeffreys' scale. The values are ad hoc, in the sense that one actually chooses at which point to reject a model in favor of another one, not unlike the traditional $68\%$ and $95\%$ confidence limits.

Also, the fact that the Bayes factor is sensitive to the priors is well known, see Ref.~ \cite{kasraft}, so we are just making an observation in order to clarify this issue. Furthermore, this can be considered as the weak statement that if the Bayes factor changes significantly under reasonable variation of prior parameter assumptions, then the data will not be able to tell us much because most of the uncertainty will come from theoretical (prior) terms rather than data (likelihood) terms. In other words, it could be described as "a fallacy of the experimenter", but this effect is not taken into account in the Jeffreys's scale.

\section{Figure of Merit\label{secfom}}

In this section we will calculate the Figure of Merit (FoM) for this general model by using the usual definition, but we will also introduce a new version which is instead useful for reconstructed quantities that are a function of $x$.

The $n\sigma$ contours are defined by the constraint equation
\be
\mathcal{C}: \chi(\vec{a})^2= \chi_{\rm min}^2+\delta \chi^2 \label{constr0}
\ee
where the value of $\delta \chi^2$ depends on the number of parameters $M$ and the number $n$ of desired $\sigma$s \cite{press92}. This important parameter $\delta \chi^2$ can be found by solving \cite{press92}
\be
1 - \mathcal{Q}\left(M/2, \delta \chi^2/2\right) = \erf\left(n/\sqrt{2}\right) \label{dchi20}
\ee
for $\delta \chi^2\geq0$, where $\mathcal{Q}(a,z)$ is the regularized incomplete gamma function $\mathcal{Q}(a,z)\equiv \frac{\Gamma(a,z)}{\Gamma(z)}$ \cite{handbook}. Equation (\ref{dchi20}) can be solved for $\delta \chi^2(M,n)$ as:
\be
\delta \chi^2(M,n) =2~ \mathcal{G}\left(\frac{M}{2},1-\erf\left(\frac{n}{\sqrt{2}}\right)\right),\label{dchi21}
\ee
where $\mathcal{G}$ is the inverse $\Gamma$ regularized function\footnote{This function can be calculated in Mathematica as $\mathcal{G}(x,y)=InverseGammaRegularized[x,y]$ and works both symbolically and numerically to arbitrary precision.}.

By using Eqs.~(\ref{chi2def1}) and (\ref{transf1})-(\ref{transf3}) we can rewrite the constraint equation as follows:
\ba
\mathcal{C}: (a-a_{\rm min})_i ~F_{ij}~ (a-a_{\rm min})_j &=& \delta \chi^2 \nn \\ \sum_{i=0}^{M-1}s_i^2 &=& \delta \chi^2 \label{constr1}
\ea
which in the rotated frame describes a hyper-sphere in $M$ dimensions.

The FoM is equal to the inverse of the volume inside the space whose boundary is given by Eq.~(\ref{constr0}) or in other words the constrained integral \ba \textrm{Vol(M)} &=& \int_{\mathcal{C}} d^M a_i \nn\\ \textrm{FoM} &=& \textrm{Vol(M)}^{-1} \label{fom1}\ea This is clearly done so that a smaller volume (better constraints) gives a higher FoM. From Eqs.~(\ref{constr1}) and (\ref{transf1})-(\ref{transf3}) the volume can be expressed as
\ba
\textrm{Vol(M)} &=& \int_{\mathcal{C}} |D|^{-1} d^M s_i \nn  \\ &=& |F|^{-1/2} V_M(\delta \chi^2) \nn \\ &=& |F|^{-1/2} \frac{\pi^{M/2}}{\Gamma(M/2+1)} \left(\delta \chi^2\right)^{M/2} \label{vol}
\ea
where in the last line we used the fact that the volume of a hyper-sphere of ``radius" $R_M=\left(\delta \chi^2\right)^{1/2}$ in $M$ dimensions, which is equal to the constrained integral in the new basis, is $V_M(\delta \chi^2)=\frac{\pi^{M/2}}{\Gamma(M/2+1)} \left(\delta \chi^2\right)^{M/2}$. Finally,
\be
\textrm{FoM}(M)= |F|^{1/2} \frac{\Gamma(M/2+1)}{\pi^{M/2}} \left(\delta \chi^2\right)^{-M/2} \label{fom2}
\ee

\begin{figure}[t!]
\centering
\vspace{0cm}\rotatebox{0}{\vspace{0cm}\hspace{0cm}\resizebox{0.95\textwidth}{!}{\includegraphics{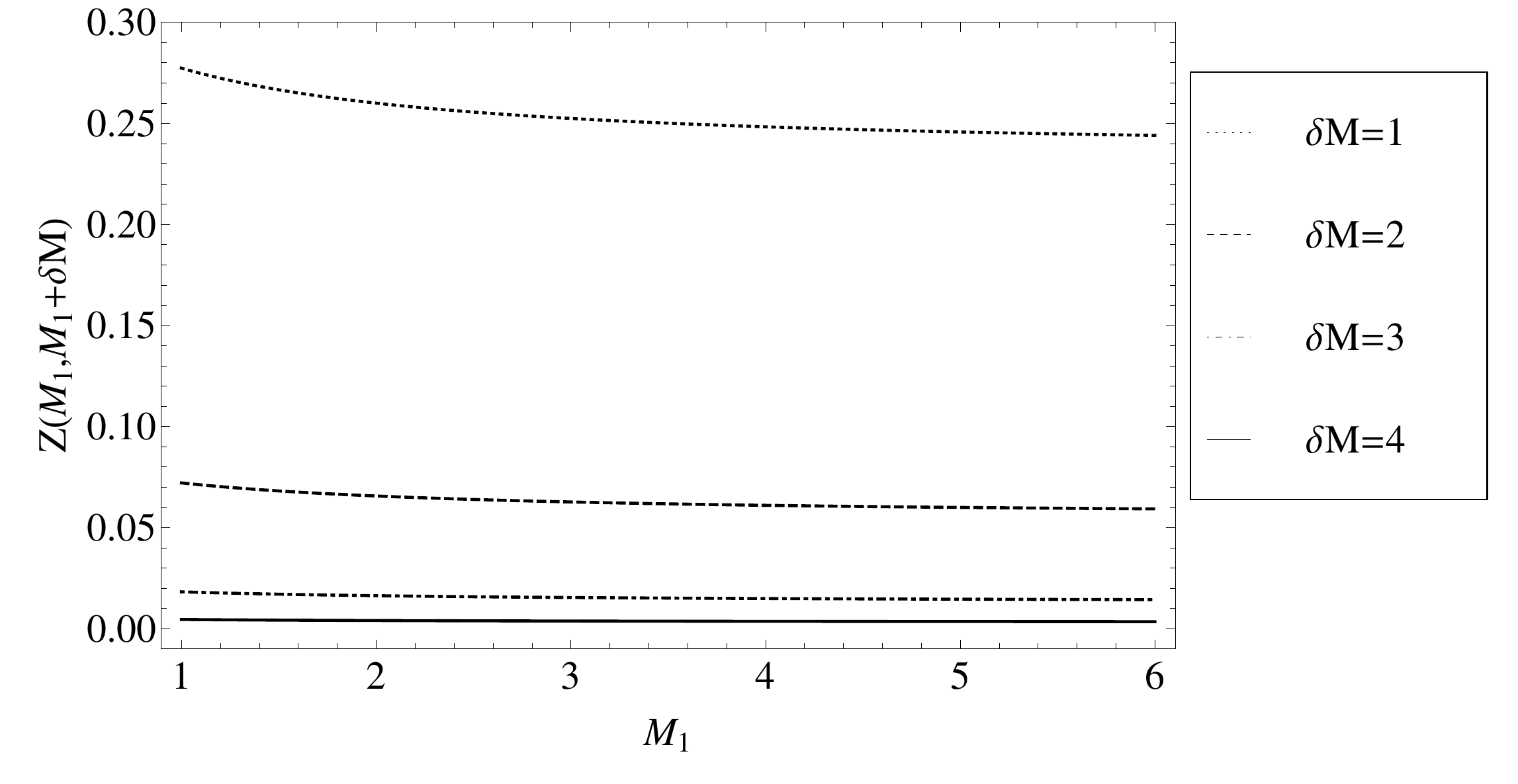}}}
\caption{The dependence of the function $Z(M,\delta M)$ given by Eq.~(\ref{Zfom}) on $M$ and $\delta M$ for $n=1$ ($1\sigma$). \label{fomZ}}
\end{figure}

\begin{figure*}[t!]
\centering
\vspace{0cm}\rotatebox{0}{\vspace{0cm}\hspace{0cm}\resizebox{0.65\textwidth}{!}{\includegraphics{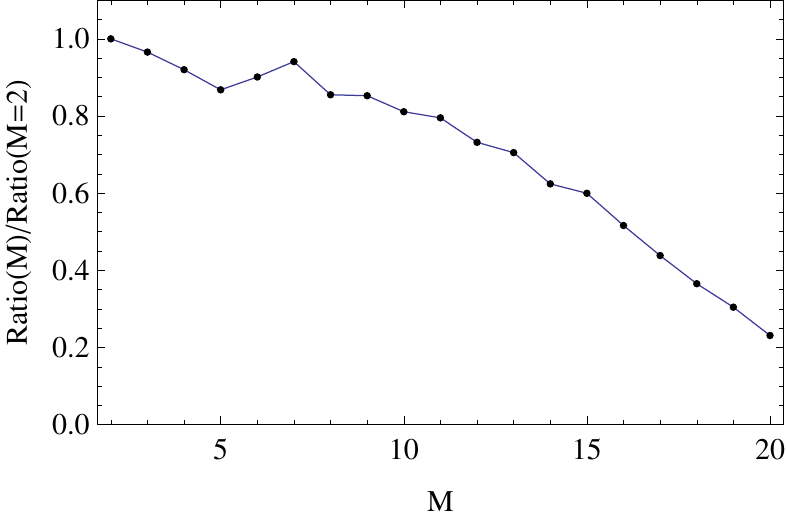}}}
\caption{The dependence of $\textrm{Ratio}(M)$ on $M$, but normalized to $\textrm{Ratio}(M=2)$. Clearly, adding more parameters does not improve the FoM, when the two models differ just by one parameter, ie $M_1-M_2=1$. \label{plotfom}}
\end{figure*}

Now, suppose we want to compare two models that have $M_1=M+\delta M$ and $M_2=M$ parameters. Then, the ratio of the FoM for the two models can be written as
\ba
\frac{\textrm{FoM}(M_1)}{\textrm{FoM}(M_2)} &=& \frac{|F^{(1)}|^{1/2}}{|F^{(2)}|^{1/2}} \left(\frac{\frac{\Gamma((M+\delta M)/2+1)}{\pi^{(M+\delta M)/2}} \left(\delta \chi^2\right)^{-(M+\delta M)/2}}{\frac{\Gamma(M/2+1)}{\pi^{M/2}} \left(\delta \chi^2\right)^{-M/2}}\right) \nn \\ &=& Z(M,\delta M)~\frac{|F^{(1)}|^{1/2}}{|F^{(2)}|^{1/2}}
\ea
where we have defined
\be
Z(M,\delta M)\equiv \frac{\frac{\Gamma((M+\delta M)/2+1)}{\pi^{(M+\delta M)/2}} \left(\delta \chi^2(M+\delta M,n)\right)^{-(M+\delta M)/2}}{\frac{\Gamma(M/2+1)}{\pi^{M/2}} \left(\delta \chi^2(M,n)\right)^{-M/2}}\label{Zfom}
\ee
The dependence of the function $Z(M,\delta M)$ on $M$ and $\delta M$ for $n=1$ ($1\sigma$) is shown in Fig. \ref{fomZ}. As it can easily be seen from Eq.~(\ref{Zfom}), $Z(M,\delta M)$  is completely independent of the data, depending solely on the number of parameters and the number $n$ of $\sigma$s. If we want to study the dependence of the FoM of the number of parameters in the case of a nested model, eg when $M_1=M+1$ and $M_2=M$, then it is convenient to define the function
\be
\textrm{Ratio}(M)\equiv \frac{\textrm{FoM}(M+1)}{\textrm{FoM}(M)} \label{ratio}
\ee
In Fig.~\ref{plotfom} we show the dependence of $\textrm{Ratio}(M)$ on $M$, but normalized to $\textrm{Ratio}(M=2)$. Clearly, adding more parameters does not improve the FoM, especially when the two models differ just by one parameter, ie $M_1-M_2=1$.

We can also explore the dependence of the FoM on the different bases $X_n(x)$, as shown in Section \ref{CoB}, but now with the same number of parameters $M$. If we denote the FoM for basis 1 as $\textrm{FoM}_1(M)$ and the FoM for basis 2 as $\textrm{FoM}_2(M)$, then by using Eqs.~(\ref{fom2} and (\ref{LFij}) we have
\be
\frac{\textrm{FoM}_1(M)}{\textrm{FoM}_2(M)}=\frac{|F_1|^{1/2}}{|F_2|^{1/2}}=|\Lambda|
\ee
where we have assumed that $F_1=\Lambda^T~F_2~\Lambda$. In the case of our example we have that $x_{max}=1.55$, so in Table~\ref{table01} we show the ratio of the FoM between the different combinations of bases. Clearly, the Chebyshev polynomials provide the best constraints out of all three cases.

\begin{table}
\begin{center}
\caption{The determinant of the transformation matrix $\Lambda_{kn}$ for various combinations of polynomials including the monomials $x^n$, the Legendre polynomials $P_n(x)$ and the Chebyshev polynomials $T_n(x)$. Clearly, the Chebyshev polynomials provide the best constraints out of all three cases. \label{table01}}
\begin{tabular}{c|ccc}
 $\frac{\textrm{FoM}_1(M)}{\textrm{FoM}_2(M)}=|\Lambda|$ & Monomials & Legendre & Chebyshev \\
\hline
 Monomials & 1.000 & 0.296 & 0.222 \\
 Legendre & 3.384 & 1.000 & 0.750 \\
 Chebyshev & 4.511 & 1.250 & 1.000\\
\end{tabular}
\end{center}
\end{table}

\section{Conclusions\label{conclusions}}

We are entering an era where progress in cosmology is driven by data, and alternative models will have to be compared and ruled out according to some consistent criterium. The most conservative and widely used approach is Bayesian model comparison. Naively, one expects the Bayes factor to act as a discriminant among competing models by penalizing those with a larger set of parameters. This has been the common use of Bayesian model comparison in cosmology in the last decade. However, by explicitly computing the Bayes factors for models that are linear with respect to their parameters, we have shown that more information is needed in order to discriminate among models. In particular, we have seen that the thresholds associated to the so called Jeffreys' scale are not as conclusive as most people think they are. We have determined how accurate its predictions are in a simple case where we fully understand and can calculate everything analytically.

We presented our results on a test of the  Jeffreys' scale, which is used for comparing models, by explicitly calculating the best fit parameters, the minimum $\chi^2$ and the Bayes factors for all models that are linear with respect to their parameters regardless of the basis used, like in Eq.~(\ref{model}). The calculation of the Bayes factor was done for both flat and gaussian priors and analytic formulas were derived in both cases. We also considered the case of changing basis by a transformation from the original \textit{polynomial} basis $X(x)$ with a set of $M$ linearly dependent parameters $a_i$ to a new set with the same number of linearly dependent parameters $M$ and a \textit{polynomial} basis $\tilde{X}(x)$. Furthermore, we also discussed the case of nested models, eg one with $M_1$ and another with $M_2\supset M_1$ parameters and we derived analytic expressions for the Bayes factor, while in Section \ref{secfom} we discussed the same problem for the Figure of Merit.

Our main result for the logarithmic Bayes factor between two models that linearly depend on the parameters, Eq.~(\ref{Bayesfac}), does not only contain the difference between the $\chi^2_{min}$ of the two models but also contains information on their covariances. Unsurprisingly, the covariances depend strongly on the data and the model at hand, thus introducing a further complication in model comparison. Therefore, the Bayes factor cannot be the only discriminant among models and thus cannot be taken as a quantitative version of Occam's razor, which simply penalizes models with a larger number of parameters. It is model predictiveness, not model simplicity, which is "rewarded" in Bayesian model comparison \cite{March:2010ex},\cite{Kunz:2006mc}. We have shown this by studying analytically the Bayes factors for models with both a small and a large number of parameters to be constrained by the same mock data. In particular, we found that  models with $M_1=4$ and $M_2=14$ parameters had similar Bayes factor ($\ln B_{12} ~ 1$), and thus were ``undecided" by Jeffreys' scale. This could only be understood if the extra 10 parameters do not contribute to the improvement of the model as a description of the data, irrespective of the fact that none of them, $M_1$ nor $M_2$, are the ``true" model. In fact, even though the one with 14 parameters gets a better $\chi^2$, of course. However, this information is not contained in the Jeffrey's scale, and one could thus be fooled by the Bayes factor to assign similar probabilities to both models.

Another similar example can be seen for $M_1=4$, $M_2=10$ and $n=7$, see Fig.~\ref{plots2} on the right, where the Bayes factor is $\log{B_{12}=0}$ meaning that these two models are totally equivalent! This simple example clearly demonstrates that the Jeffrey's scale is an inadequate tool for model comparison, since it completely fails even in this simple example. Also, as it can be seen in the two plots of Fig.~\ref{plots2} the results and the conclusions for the two models $M_1$ and $M_2$ are very sensitive in the choice of the priors $\Delta a_i$.

Furthermore, we tested the Jeffrey's scale when the real model was an exponential but also a simple low order polynomials so that the comparison with the polynomials of degree $M$ is more fair. As it can be seen in Fig 5 which is based on the low order polynomial (note that the yellow and red regions of the contour plots correspond to cases where $B_{ij}>150$ or $\ln(B_{ij})>5$) the point $(M_1, M_2)=(4,14)$ has $\ln(B_{ij})\sim 5.5$ which means that according to the Jeffrey's scale $M_2$ should have been ruled out, but a slightly different choice of models $(M_1, M_2)=(4,12)$, has $\ln(B_{ij})\sim 1.1$ which means that these two are equally likely when clearly $M_2$ is should again be ruled out. So, this proves beyond reasonable doubt that the threshold, eg $\ln(B_{ij})>5$ where one chooses to reject a model is completely arbitrary thus making the Jeffreys' scale neither a robust nor a reliable tool for model comparison.

To conclude, the Jeffreys' scale seems to be susceptible to two types of errors, the so called ``Type I" and ``Type II" errors. Regarding the "Type I error" or false positive, ie an incorrect rejection of a true hypothesis: in Fig. \ref{plotsM12vsreal}, the real exponential model is rejected in favor of the simple polynomials of order up to $M \sim 12$. Regarding the "Type II error" or false negative, ie a failure to reject a false hypothesis: consider the case $(M_1,M_2)=(4,12)$ in Fig. \ref{plots2other} that has $\ln(B_{ij})\sim 1.1$, which means that these two are equally likely when clearly $M_2$ should have been ruled out. While the Bayes factors are clearly related to the probabilities that one of the two models are more likely than the other, the threshold values of Table \ref{table1} of the Jeffreys' scale used to reject a model in favor of another, are open to interpretation. To make it more clear, the problem is not with the probabilistic interpretation of the Bayes factor, but with the Jeffreys' scale itself. The latter, just represents the threshold after which one is forced to reject a model, usually when $\log{B_{12}}>5$ (i.e. strong evidence). What we have found is that, even when one would expect that a model with 14 parameters would and should be ruled out with respect to one with 4 parameters, it was allowed according to the Jeffreys' scale, since $\log{B_{12}}\sim1$!

Obviously, having an {\em ad hoc} scale for model comparison where the thresholds are the same irrespectively of the models and the data, used as a ``one size fits  all" tool, can lead to \textit{biased} conclusions. The situation can potentially be even worse in cases where the models are not as simple as in the one at hand, e.g. consider the case in cosmology where the models used are non-linear and substantially more complicated \cite{Tsujikawa:2010sc}.

\section*{Acknowledgements}
We would like to thank A.~Liddle and R.~Trotta for very useful and enlightening discussions. Also, we would like to thank an anonymous referee for his/her valuable input and corrections to our paper. We acknowledge financial support from the Madrid Regional Government (CAM) under the program HEPHACOS S2009/ESP-1473-02, from MICINN under grant AYA2009-13936-C06-06 and Consolider-Ingenio 2010 PAU (CSD2007-00060), as well as from the European Union Marie Curie Initial Training Network UNILHC PITN-GA-2009-237920. S.~N. is supported by CAM through a HEPHACOS Fellowship.

\appendix
\section{General linear least-squares fitting\label{fitting}}

\subsection{Minimization}

In this section we will briefly discuss the case of the general linear least squares fitting. Given some data that consist of $N$ measurements $(x_i,y_i,\sigma_i)$, where $i=(1,2,...,N)$, and a model which is a linear combination of $M$ functions,
\be \label{model}
y(x)= \sum_{i=0}^{M-1} a_i X_i(x)\,,
\ee
then the fitness of the model with respect to the data and the parameters $a_i$, for $i=(0,1,...,M-1)$,
can be found by calculating the $\chi^2(\vec{a})$ defined as \cite{press92}, \cite{Young:2012kg}
\ba
\chi^2(\vec{a})&\equiv& \sum_{i=1}^N \left(\frac{y_i-y(x_i;a_j)}{\sigma_i}\right)^2 \nn \\
&=& \sum_{i=1}^N \left(\frac{y_i-\sum_{j=0}^{M-1}a_j X_j(x_i)}{\sigma_i}\right)^2 \label{chi2def}
\ea

The base $X_i(x)$ can be any set of $M$ functions, e.g. monomials like $\{x^i\}_{i=0}^{M-1}$ or Chebyshev polynomials $\{T_i(x)\}_{i=0}^{M-1}$, of order $M$. The latter are a set of orthogonal polynomials that can act as a base of functions with the property that when $x\in[-1,1]$ they have the smallest maximum deviation from the true function at any given order $M$. The first few Chebyshev polynomials are $T_0(x)=1,~T_1(x)=x,~T_2(x)=-1+2 x^2,~T_3(x)=-3 x+4 x^3$. When $x\in[-1,1]$, the variable $x$ can be written as $x=\cos(\theta)$ and the polynomials can also be expressed as $T_n(\cos(\theta))=\cos(n\theta)=\cos(n\arccos(x))$, which implies that $|T_n(x)|\leq1$. Since in general our data will not be in the range $[-1,1]$, we can normalize $x$ by using $\tilde{x}=\frac{2x}{x_{max}}-1$ and using instead the basis $T_n\left(\tilde{x}\right)\equiv T_n(\frac{2x}{x_{max}}-1)$, where $x_{max}$ is the maximum value of the $N$ data $x_i$. From now on we will assume that $x$ has been normalized and we will drop the tilde on $x$. Finally, we will mostly follow the notation of Ref. \cite{press92}.

The best-fit parameters of the model can be found by minimizing the $\chi^2(\vec{a})$ of Eq. (\ref{chi2def}) with respect to the parameters $a_j$. This is done by demanding that the derivatives of $\chi^2(\vec{a})$ are equal to zero at the minimum, i.e. $ \partial_j \chi^2(a_k)=0$. Then, this gives \cite{press92}, \cite{Young:2012kg}

\ba
\partial_j \chi^2(a_k) &=& \sum_{i=1}^N \left(\sum_{k=0}^{M-1} (-2)\frac{\partial a_k}{\partial a_j}\frac{X_k(x_i)}{\sigma_i}\right)\left(\frac{y_i-\sum_{m=0}^{M-1}a_m X_m(x_i))}{\sigma_i}\right) \nn \\
&=& (-2)\sum_{i=1}^N \frac{X_j(x_i)}{\sigma_i} \left(\frac{y_i-\sum_{m=0}^{M-1}a_m X_m(x_i)}{\sigma_i}\right) = 0 \nn
\ea
or equivalently,
\ba
\left(\sum_{i=1}^N \frac{y_i X_j(x_i)}{\sigma_i^2}\right) &=&\left(\sum_{k=0}^{M-1}\sum_{i=1}^N \frac{X_j(x_i) X_k(x_i)}{\sigma_i^2} a_k \right)\,.
\label{system}
\ea
If we define the Fisher Matrix $F_{ij}$ and the constant vector $\beta_j$ as
\ba
F_{ij}&=&\frac{1}{2} \partial_{ij} \chi^2|_{\rm min} = \sum_{k=1}^N \frac{X_i(x_k) X_j(x_k)}{\sigma_k^2} =\textrm{ const.}\\
\beta_j &=& \sum_{i=1}^N \frac{y_i X_j(x_i)}{\sigma_i^2}=\textrm{ const.},
\ea
where as usual $j=(0,1,...,M-1)$, then Equation (\ref{system}) can be rewritten in matrix form and easily solved for the best-fit parameters $\vec{a}_{\rm min}$ as
\ba
\beta_j&=&F_{jk}~a_{k, {\rm min}} \nn \\
a_{k, {\rm min}}&=& F^{-1}_{kj} \beta_j =C_{kj}\beta_j ,\label{linsol}
\ea
where $C_{kj} \equiv F^{-1}_{kj}$ is the covariance matrix. If we define the parameter $S_{y}\equiv \sum_{i=1}^N y_i^2/\sigma_i^2$, then the $\chi^2$ at the minimum can be written as
\ba
\chi^2_{\rm min}&=&S_{y}-C_{ij}\beta_i\beta_j \nn \\ &=&S_{y}-F_{ij}a_{i, {\rm min}}a_{j, {\rm min}}\,.
\label{bestchi2}
\ea
Finally, the $1\sigma$ errors on the best-fit parameters are given by the diagonal elements of the covariance matrix
\be
\sigma(a_k)^2=C_{kk}
\ee
At this point we should note that by doing a Taylor expansion around the minimum the $\chi^2(\vec{a})$ can also be written as
\be
\chi^2(\vec{a})= \chi^2_{\rm min}+ (a-a_{\rm min})_i ~F_{ij}~ (a-a_{\rm min})_j,
\label{chi2def1}
\ee
since the first derivatives at the minimum are by definition equal to zero and all higher derivatives $\partial^n_{i_1...i_n} \chi^2(\vec{a})$ for $n\geq3$ vanish like in our model.

\subsection{Change of basis\label{CoB}}
At this point we should note that the functional form of the results of Eqs.~(\ref{linsol})-(\ref{chi2def1}) is completely independent of the basis used, regardless of it being some combination of polynomials (monomials or Chebyshev polynomials) or something more complicated, e.g. $\sin(n~x)$ etc. For example, in the case of the monomials the Fisher matrix is equal to
\be
F_{ij}=\frac{1}{2} \partial_{ij} \chi^2|_{\rm min} = \sum_{k=1}^N \frac{x_k^i x_k^j}{\sigma_k^2},
\ee
for $i=(0,1,...,M-1)$. If at this point we define a constant
\be
S_n\equiv \sum_{k=1}^N\frac{x_k^n}{\sigma_k^2},
\ee
where for example $S_0=\sum_{k=1}^N\frac{1}{\sigma_k^2}$, $S_1=\sum_{k=1}^N\frac{x_k}{\sigma_k^2}$, $S_2=\sum_{k=1}^N\frac{x_k^2}{\sigma_k^2}$ and so on, then in the case that $M=3$ the Fisher matrix will be given by
\be F_{ij}=
  \left(
\begin{array}{ccc}
 S_0 & S_1 & S_2 \\
 S_1 & S_2 & S_3 \\
 S_2 & S_3 & S_4
\end{array}
\right),
\ee where the constants $S_n$ will only depend on the data.

For the Chebyshev polynomials the Fisher matrix is equal to
\be
F_{ij}=\frac{1}{2} \partial_{ij} \chi^2|_{\rm min} = \sum_{k=1}^N \frac{T_i(x_k) T_j(x_k)}{\sigma_k^2}.
\ee
For example, in the case that $M=3$ the Fisher matrix will be given by
\be F_{ij}=
  \left(
\begin{array}{ccc}
 \sum_{k=0}^N \frac{T_0(x_k)^2}{\sigma_k^2} & \sum_{k=0}^N \frac{T_0(x_k)T_1(x_k)}{\sigma_k^2} & \sum_{k=0}^N \frac{T_0(x_k)T_2(x_k)}{\sigma_k^2} \\
 \sum_{k=0}^N \frac{T_0(x_k)T_1(x_k)}{\sigma_k^2} & \sum_{k=0}^N \frac{T_1(x_k)^2}{\sigma_k^2} & \sum_{k=0}^N \frac{T_1(x_k)T_2(x_k)}{\sigma_k^2} \\
 \sum_{k=0}^N \frac{T_0(x_k)T_2(x_k)}{\sigma_k^2} & \sum_{k=0}^N \frac{T_1(x_k)T_2(x_k)}{\sigma_k^2} & \sum_{k=0}^N \frac{T_2(x_k)^2}{\sigma_k^2}
\end{array}
\right),
\ee which obviously is constant and will only depend on the data at hand\footnote{At this point we should remind the reader of our shorthand convention that $x$ is normalized, so for example by writing $T_0(x_k)$ we actually imply $T_0\left(\frac{2x_k}{x_{max}}-1\right)$.}. Similar expressions can be derived for other cases as well.

In general, if we change basis from $X_i(x)$ to some new \textit{polynomial} basis $\tilde{X}_i(x)$ of the same \textit{order}, then we will have to replace the parameters $a_k$ with some new parameters $\tilde{a}_k$, assuming that we still have the same number of linearly dependent parameters $M$. However, the function $y(x)$ will still be the same, so

\ba \label{modelnb}
y(x)&=& \sum_{i=0}^{M-1} a_i X_i(x)\nn \\
&=&\sum_{i=0}^{M-1} \tilde{a}_i \tilde{X}_i(x)\,.
\ea
Let one of the two bases satisfy an orthogonality relation with a weight $w(x)$:
\be
\int_{x_1}^{x_2} dx ~w(x)~\tilde{X}_i(x) \tilde{X}_j(x) =c_i \delta_{ij}
\ee
for some constants $c_i$. Then, by using Eq.~(\ref{modelnb}) and the orthogonality relation we can derive a transformation $\Lambda$ between the two sets of parameters and for the other quantities of interest:
\ba
a_i&=& \Lambda^{-1}_{ij} \tilde{a}_j  \\
\beta_i &=& \Lambda_{ij}^T \tilde{\beta}_j  \\
F_{ij}&=& \Lambda_{ik}^T \tilde{F}_{kl} \Lambda_{lj} \label{LFij} \\
\tilde{C}_{ij}&=& \Lambda_{ik} C_{kl} \Lambda_{lj}^T
\ea
Finally,  as it can easily be seen from Eq.~(\ref{bestchi2}) the $\chi^2_{min}$ as expected  is invariant under the transformation $\Lambda$.

For example, we will now consider the change of basis from the monomials to the Chebyshev polynomials. In this case we will have
\be \label{modelnb1}
\sum_{n=0}^{M-1} a_n x^n= \sum_{n=0}^{M-1} \tilde{a}_n T_n\left(\frac{2x}{x_{max}}-1\right)\,,
\ee
Remembering the fact that the Chebyshev polynomials satisfy the orthogonality relation
\be
\int_{-1}^{1}dz \frac{T_k(z)T_n(z)}{\sqrt{1-z^2}}=k_n\delta_{nk}
\ee
where $k_0=\pi$ and $k_n=\pi/2$ for $n\geq1$. Then we can multiply Eq.~(\ref{modelnb1}) with the appropriate factors $\frac{T_n(z)}{\sqrt{1-z^2}}$, where $z=\frac{2x}{x_{max}}-1$ and by integrating both sides over $z\in[-1,1]$ we get the transformation between the two sets of parameters as
\be
\tilde{a}_n = \Lambda_{kn} a_k
\ee
where the constant matrix $\Lambda$ is given by
\be
\Lambda_{kn}=k_n^{-1} \int_{-1}^{1}dz \frac{\left(\frac{x_{max}}{2}(z+1)\right)^k T_n(z)}{\sqrt{1-z^2}}
\ee

In Table~\ref{table0} we show the determinant of the transformation matrix $\Lambda_{kn}$ for various combinations of polynomials including the monomials $x^n$, the Legendre polynomials $P_n(x)$ and the Chebyshev polynomials $T_n(x)$. These values are particularly important in the estimation of the Figure of Merit, as shown in a later section. Finally, we should stress that these results are only valid for a transformation from the original \textit{polynomial} basis $X(x)$ with a set of $M$ linearly dependent parameters $a_i$ to a new set with the same number of linearly dependent parameters $M$ and a \textit{polynomial} basis $\tilde{X}(x)$.

\begin{table}
\begin{center}
\caption{The determinant of the transformation matrix $\Lambda_{kn}$ for various combinations of polynomials including the monomials $x^n$, the Legendre polynomials $P_n(x)$ and the Chebyshev polynomials $T_n(x)$. These values are particularly important in the estimation of the Figure of Merit, as shown in a later section. \label{table0}}
\begin{tabular}{c|ccc}
\hspace{2pt} $|\Lambda|$\hspace{2pt} &\hspace{2pt} Monomials \hspace{2pt}& \hspace{2pt}Legendre \hspace{2pt}& \hspace{2pt}Chebyshev\hspace{2pt}\\
\hline
\hspace{2pt}Monomials \hspace{2pt}&\hspace{2pt} 1 \hspace{2pt}& \hspace{2pt}$\frac{x_m^3}{12}$ & \hspace{2pt}$\frac{x_m^3}{16}$ \hspace{2pt}\\
\hspace{2pt}Legendre \hspace{2pt}&\hspace{2pt} \hspace{2pt}$\hspace{2pt}\frac{12}{x_m^3}$ \hspace{2pt}& \hspace{2pt}1\hspace{2pt} &\hspace{2pt} $\frac{3}{4}$ \hspace{2pt}\\
\hspace{2pt} Chebyshev \hspace{2pt}&\hspace{2pt} \hspace{2pt}$\hspace{2pt}\frac{16}{x_m^3}$ \hspace{2pt}& \hspace{2pt}$\frac{4}{3}$ \hspace{2pt}& 1
\end{tabular}
\end{center}
\end{table}

\subsection{Likelihood calculations}

After having determined the best-fit parameters in the previous section, we will now define the likelihood for our model. This is defined as \cite{press92}:
\be
\mathcal{L}=\mathcal{N}\exp\left(-\chi^2(\vec{a})/2\right), \label{likelihood}
\ee
where the parameter $\mathcal{N}$ can be found by normalizing the likelihood, ie integrating it over all parameters. In our case this means :
\ba
\int \mathcal{L}d\vec{a}= \int_{-\infty}^{\infty} \mathcal{N}\exp\left(-\chi^2(\vec{a})/2\right) d\vec{a} &=&\nn \\ \mathcal{N} e^{-\chi^2_{\rm min}/2} \int _{-\infty}^{\infty} e^{-1/2(a-a_{\rm min})_i ~F_{ij}~ (a-a_{\rm min})_j} da_0 da_1...da_{M-1} &=&1 \label{int1}
\ea

To proceed we now have to rotate the parameters to a basis where they are not correlated with each other. To do so we define a new variable $s_i\equiv D_{ij}\left(a_j-a_{j, {\rm min}}\right)$, where $D_{ij}$ can be found by decomposing the inverse covariance matrix $F=C^{-1}=D^T D$ by using Cholesky decomposition\footnote{Cholesky decomposition can easily be implemented in computer programs such as Mathematica. For example, in the latter the Cholesky decomposition of a matrix $M=D^{T} D$ is given by $D=CholeskyDecomposition[M]$. This works both symbolically and numerically.}. Then, we have that $ds_1 ... ds_N=\left\vert D \right\vert df_1...df_N$ and the integration can proceed as usual and the normalization can be found. Going to the new basis we have
\ba
s_i &\equiv& D_{ij}\left(a_j-a_{j, {\rm min}}\right)\label{transf1}\\ds_1 ... ds_N&=&\left\vert D \right\vert da_0 da_1...da_{M-1} \\ \left\vert D \right\vert &=& \left\vert F \right\vert^{1/2}=\left\vert C \right\vert^{-1/2} \label{transf3}
\ea
where $a_{j, {\rm min}}$ is to be understood as the value of the $j$th parameter $a_j$ at its best-fit value (the ``minimum"). Then, Eq.~(\ref{int1}) becomes
\ba
\mathcal{N}~e^{-\chi^2_{\rm min}/2} \int_{-\infty}^{+\infty} e^{-\sum_{i=0}^{M-1} s_i^2/2} \left\vert D \right\vert^{-1} \prod_{i=0}^{M-1} ds_i &=& \nn \\ \mathcal{N}~e^{-\chi^2_{\rm min}/2} \left\vert F \right\vert^{-1/2} \prod_{i=0}^{M-1} \int_{-\infty}^{+\infty} e^{-s_i^2/2} ds_i &=& \nn \\ \mathcal{N}~e^{-\chi^2_{\rm min}/2} \left\vert F \right\vert^{-1/2} (2 \pi)^{M/2} &=&1, \nn \\
\ea
and finally,
\be
\mathcal{N}=e^{\chi^2_{\rm min}/2} \left\vert F \right\vert^{1/2} (2 \pi)^{-M/2}. \label{norm} \ee
Unsurprisingly, the resulting normalized likelihood now becomes
\ba
\mathcal{L}&=&\frac{\left\vert F \right\vert^{1/2}}{(2\pi)^{M/2}} \exp\left(-(\chi^2(\vec{a})-\chi^2_{\rm min})/2\right), \label{likelihood1} \\ &=&\frac{1}{(2\pi)^{M/2} \left\vert C \right\vert^{1/2}} \exp\left(-(\chi^2(\vec{a})-\chi^2_{\rm min})/2\right), \nn
\ea
where in the last line we used the fact that $\left\vert F \right\vert =\left\vert C \right\vert^{-1}$.

\end{document}